\documentclass[12pt]{article}
\usepackage{amssymb}
\usepackage{amsfonts}
\usepackage{mathrsfs}
\usepackage{amsmath}
\usepackage{slashed}
\usepackage{url}
\usepackage{bm}
\usepackage{upgreek}
\usepackage{hyperref}
\usepackage{graphicx}
\usepackage{setspace}
\usepackage[compact]{titlesec}
\usepackage[letterpaper,left=1.25in,right=1.25in,top=1in,bottom=1in,nohead]{geometry}
\usepackage{cite}
\usepackage{abstract}
\usepackage{lmodern}
\usepackage[T1]{fontenc}
\usepackage{xcolor}
\definecolor{dark-blue}{rgb}{0.15,0.15,0.4}

\hypersetup{
    colorlinks=true,      
    linkcolor=black,         
    citecolor=black,       
    urlcolor=dark-blue          
}

\titleformat{\section}{\normalsize\bfseries}{\thesection}{1em}{}
\titleformat{\subsection}{\normalsize\bfseries}{\thesubsection}{1em}{}

\numberwithin{equation}{section}

\def\^{{\wedge}}
\def\*{{\star}}
\def\bar{\overline}

\def\e#1{{\rm e}^{\, #1}}
\def\wt#1{\widetilde#1}

\def\Lie{\mathop{\rm Lie}}
\def\Vol{\mathop{\rm Vol}}
\def\Map{{\mathop{\rm Map}}}

\def\Ker{{\mathop{\rm Ker}}}
\def\Im{\mathop{\rm Im}}

\def\mod{\mathop{\rm mod}}

\def\Tor{{\mathop{\rm Tor}}}

\def\vol{{\mathop{\rm vol}}}
\def\Vol{{\mathop{\rm Vol}}}

\def\BC{{\mathbb C}}

\def\BL{{\mathbb L}}

\def\BR{{\mathbb R}}

\def\BZ{{\mathbb Z}}

\def\CA{{\mathcal A}}

\def\CD{{\mathcal D}}

\def\CF{{\mathcal F}}
\def\CG{{\mathcal G}}
\def\CH{{\mathcal H}}

\def\CO{{\mathcal O}}

\def\CX{{\mathcal X}}

\def\Fe{{\mathfrak e}}

\def\SA{{\mathsf A}}
\def\SC{{\mathsf C}}
\def\SJ{{\mathscr J}}
\def\SL{{\mathsf L}}
\def\SQ{{\mathsf Q}}
\def\SS{{\mathsf S}}
\def\SW{{\mathsf W}}
\def\SV{{\mathsf V}}

\newcommand{\version}{\normalfont May 2014}

\begin{document}
\begin{titlepage}
\begin{flushright}
{\small\ttfamily arXiv:1405.2123}
\end{flushright}
\begin{center}
\vspace{2cm}
{\large\bfseries Global Aspects of Abelian Duality\\[1ex] in Dimension
  Three}\\
\vspace{1cm}
Chris Beasley\\
\vspace{3mm}
{\small\sl Department of Mathematics, Northeastern University, Boston, MA 02115}
\vspace{-5mm}
\end{center}
\begin{abstract}
\baselineskip=18pt
In three dimensions, an abelian gauge field is related by duality to a
free, periodic scalar field.  Though usually considered on $\BR^3$,
this duality can be extended to a general three-manifold $M$, in
which case topological features of $M$ become important.  Here I comment
upon several of these features as related to the partition
function on $M$.  In a companion article, I discuss similarly the
algebra of operators on a surface of genus $g$.
\end{abstract}
\vfill\hskip 1cm\version
\end{titlepage}
\begin{onehalfspace}
\tableofcontents\noindent\hrulefill
\section{Introduction}\label{Intro}

In classical field theory, abelian duality amounts to a simple
relation between the solutions of certain linear partial differential equations
on $\BR^n$ for ${n\ge 2}$.  The most elementary example occurs in dimension
two, for which one considers a harmonic function ${\phi\in
C^\infty(\BR^2)}$ satisfying the Laplace equation 
\begin{equation}\label{HARMS}
\triangle\phi \,=\, d^\dagger d\phi \,=\,
0\,,\qquad\qquad d^\dagger = -\*\,d\,\*\,.
\end{equation}
Here $\*$ is the Hodge star operator determined by the Euclidean
metric on $\BR^2$.  

According to \eqref{HARMS}, the one-form $\*d\phi$ is
closed.  Since $\BR^2$ has trivial cohomology in degree one, $\*d\phi$
is also exact.  Hence one can associate to $\phi$ another function ${\psi\in
C^\infty\!\left(\BR^2\right)}$ via the duality relation 
\begin{equation}\label{ABTWO}
\*d\phi \,=\, d\psi\,.\end{equation}
The relation in \eqref{ABTWO} determines the function $\psi$ up to the
addition of a constant, and $\psi$ is automatically harmonic by virtue
of the identity
\begin{equation}\label{HARMP}
\triangle\psi \,=\, d^\dagger d\psi \,=\, \* d^2\phi
\,=\, 0\,.
\end{equation}
The classical duality in \eqref{ABTWO} thus relates one solution of the
Laplace equation on $\BR^2$ to another, distinct solution.  As
well-known, the dual harmonic functions $\phi$ and $\psi$ can be
combined as the real and imaginary parts of a single holomorphic
function on ${\BR^2 \simeq \BC}$.

Similarly in dimension four, if $A$ is a $U(1)$-connection on $\BR^4$
which solves the source-free Maxwell equation 
\begin{equation}\label{MAXWELL}
d\*F_A \,=\, 0\,,\qquad\qquad\qquad F_A \,=\, dA\,,
\end{equation}
then $A$ determines another $U(1)$-connection $B$ up to gauge
equivalence via 
\begin{equation}\label{MAXWELLD}
\*F_A \,=\, F_B\,.
\end{equation}
By virtue of the Bianchi identity ${dF_A = 0}$, the curvature of the
connection $B$ also satisfies ${d\*F_B = 0}$, so the classical
electric-magnetic duality in \eqref{MAXWELLD} relates distinct solutions of
the source-free Maxwell equation on $\BR^4$.

In this paper we are concerned with abelian duality in dimension
three.  In that case, classical abelian duality relates a $U(1)$-connection $A$
which solves the Maxwell equation on $\BR^3$ to a harmonic function
$\phi$ on $\BR^3$.  By analogy to dimensions two 
and four, the classical duality relation in dimension three is 
\begin{equation}\label{DUALTD} 
\*F_A \,=\, e^2\,d\phi\,.
\end{equation}
Unlike the preceding duality relations, the classical 
duality relation on $\BR^3$ involves the electric coupling
$e^2$, which appears in the Maxwell action 
\begin{align}\label{SMAX}
{\bf I}(A) \,&=\, \frac{1}{4\pi e^2} \int_{\BR^3} F_A
\^\* F_A\,,\cr
&=\, \frac{1}{8\pi e^2} \int_{\BR^3} d^3x\,\sqrt{g} \,\, F_{A, m
n}^{} \, F_A^{\,\,\,m n}\,,\qquad\qquad m,n=1,2,3\,.
\end{align}
The factor of ${1/4\pi}$ in the first line of \eqref{SMAX} is simply a
notational convenience, which will eliminate other factors later. 
In the second line of \eqref{SMAX}, we rewrite the Maxwell action in
components with respect to an arbitrary Riemannian metric $g$ on
$\BR^3$, and we follow the standard Einstein convention in summing
over the repeated indices $m$ and $n$ on the curvature $F_A$ of the
gauge field.

The appearance of the Maxwell coupling $e^2$ in the duality relation
\eqref{DUALTD} is the first of several peculiarities specific to
dimension three.  Most famously, the duality relations in two and four
dimensions are invariant under conformal transformations, which preserve
both the one-form $\*d\phi$ on $\BR^2$ in \eqref{ABTWO} and the
two-form $\*F_A$ on $\BR^4$ in \eqref{MAXWELLD}.  But in dimension
three, if the metric $g$ is scaled by a constant factor  
\begin{equation}\label{SCALE}
g \,\longmapsto\, \Lambda^2\,g\,,\qquad\qquad\qquad \Lambda
\,\in\,\BR_+\,,
\end{equation}
the dual one-form $\*F_A$ in \eqref{DUALTD} scales non-trivially as well,
\begin{equation}\label{SCALEII}
\*F_A \,\longmapsto\,\Lambda^{-1}\,\*F_A\,.
\end{equation}
In components, $\*F_A$ is given by ${\sqrt{g}\,\epsilon_{m n p}\,F^{m 
    n}\,dx^p}$, where $\epsilon_{m n p}$ is the anti-symmetric tensor
on three indices, with fixed normalization ${\epsilon_{1 2 3}=+1}$.  The
scaling for $\*F_A$ in \eqref{SCALEII} follows from the combined scalings of
$\sqrt{g}$ with weight $\Lambda^3$ and $F^{m n}$ with weight $\Lambda^{-4}$.
As a result, neither the duality relation in \eqref{DUALTD} nor the
classical Maxwell action in \eqref{SMAX} is invariant under scale,
much less conformal, transformations of the metric on $\BR^3$.

On the other hand, if the transformation of the metric in \eqref{SCALE} is supplemented by a
non-trivial scaling for the electric coupling $e^2$ itself,
\begin{equation}\label{SCALEIII} 
e^2 \,\longmapsto\, \Lambda^{-1} \, e^2\,,
\end{equation}
then the Maxwell action on $\BR^3$ is invariant.  Because
$\*F_A$ and $e^2$ transform with identical weights, the abelian duality 
relation in \eqref{DUALTD} is also preserved under scaling.
Conversely, the appearance of $e^2$ in the duality relation is dictated by
invariance under the transformations in \eqref{SCALE} and
\eqref{SCALEIII}.    See Section $1.2$ of \cite{Jackiw:2011vz} for
more about scale and conformal transformations vis-\`a-vis duality in
three-dimensional Maxwell theory.

Each of the classical duality relations in \eqref{ABTWO},
\eqref{MAXWELLD}, and \eqref{DUALTD} extends to an equivalence of free
quantum field theories defined on an arbitrary Riemannian manifold
$\Sigma$, $X$, or $M$ of corresponding dimension two, four, or three.
Though the relevant pairs of quantum field theories are themselves trivial, the
equivalence between them is generally non-trivial and may depend in
interesting ways on the topology of the underlying manifold. 

These topological issues are particularly sharp when $M$ is a closed
three-manifold.  In that case, the global analysis of the classical Laplace
equation on $M$ is very different from the global analysis of the classical
Maxwell equation on $M$.  Solutions to the Laplace equation will be
unique up to scale, but solutions to the Maxwell 
equation generically fall into continuous families, parametrized by
the holonomies of the gauge field.  So on a general
three-manifold, there is no hope to interpret abelian duality classically,
as a one-to-one correspondence \eqref{DUALTD} between solutions of
the Laplace and Maxwell equations on $M$.  Instead, abelian duality on $M$
must be interpreted as an inherently quantum phenomenon.

A basic observable in any quantum field theory is the partition
function, and as an initial question, one can ask
how the partition function transforms under duality.  Naively, one  might expect the
partition function to be invariant under duality, but famously in
dimensions two and four, this is not so.

The most elegant statement \cite{WittenGF} occurs for
electric-magnetic duality of Maxwell theory on a four-manifold $X$.
(See also \cite{Verlinde:1995mz} for related observations.)  In this
case, the Maxwell partition function $Z_X$ depends upon both the
Maxwell coupling $e^2$ and an angular parameter $\theta$ which enters
the classical Lagrangian through the topological pairing
\begin{equation}\label{ITHETA}
\begin{aligned}
{\bf I}_\theta(A) \,&=\, \frac{i\,\theta}{8 \pi^2} \int_X F_A\^F_A\,,\\
&=\, \frac{i\,\theta}{32 \pi^2} \int_X d^4x\, \sqrt{g} \,
\epsilon_{mnpq} \, F_A^{m n} \, F_A^{p q}\,.
\end{aligned}
\end{equation}
When $X$ is a spin-manifold, the normalization in \eqref{ITHETA}
ensures that $\theta$ has period $2\pi$.  Otherwise, $\theta$ has period $4\pi$.

The angular parameter $\theta$ naturally complexifies the electric coupling
$e^2$ via 
\begin{equation}
\tau \,=\, \frac{\theta}{2\pi} \,+\, \frac{4\pi i}{e^2}\,,
\end{equation}
and electric-magnetic duality acts upon $\tau$ as a modular transformation ${\tau
  \,\mapsto\, -1/\tau}$.  Moreover, as shown by direct computation
in \cite{WittenGF}, the Maxwell partition function $Z_X(\tau)$ on $X$
transforms under duality as a non-holomorphic modular form with
weights ${\frac{1}{4}\!\left(\chi-\sigma,\,\chi+\sigma\right)}$,
\begin{equation}\label{MODULAR}
Z_X(-1/\tau) \,=\, \tau^{\frac{1}{4}\left(\chi-\sigma\right)} \,
\bar\tau^{\frac{1}{4}\left(\chi+\sigma\right)} \,
Z_X(\tau)\,.
\end{equation}
Here $\chi$ and $\sigma$ are the respective Euler character and
signature of the four-manifold $X$.  The non-trivial transformation
law for $Z_X(\tau)$ in \eqref{MODULAR} is kind of gravitational
anomaly for duality, since both the Euler character and  signature can
be represented as the integrals of local densities constructed from the
Riemann tensor on $X$.  See also the discussion in Section
$3$ of \cite{VafaTF}, where the modular anomaly in electric-magnetic
duality was originally noted in the context of supersymmetric
Yang-Mills theory.

Similarly on a Riemann surface $\Sigma$, the classic one-loop shift
\cite{Buscher:1985,BuscherSK,BuscherQJ,Giveon:1994fu,RocekPS} in the
dilaton under T-duality represents a comparable topological effect,
depending again on the Euler character of $\Sigma$.

One motivation for the present work is to point out a modular
property roughly analogous to \eqref{MODULAR} for the partition function 
of Maxwell theory on a closed, orientable three-manifold $M$.  

Such modularity in three dimensions may sound surprising, because the
partition function on $M$ (as opposed to the partition function on
$X$) can have no interesting dependence
on the electric coupling $e^2$.  A priori, the Maxwell partition
function $Z_M(e^2, g)$ depends upon both the 
coupling $e^2$ and the Riemannian metric $g$ on $M$.  However, the
scale transformations in \eqref{SCALE} and \eqref{SCALEIII} together 
preserve $Z_M(e^2, g)$ and can always be used to set ${e^2 = 1}$, so
that any dependence on $e^2$ can be effectively absorbed
into the dependence on the metric.  In addition, the Euler character
of any closed, orientable three-manifold vanishes, and there are no
other local, generally covariant invariants of $M$ that could appear
in an anomaly such as \eqref{MODULAR}.

The situation changes, though, as soon as we include additional
parameters which play a role in three dimensions analogous to
the role of the $\theta$-angle in four dimensions.
Very briefly, in dimension three the topological parameter
${\upzeta\in\CH^1_\BC(M)}$ will be a complex harmonic one-form, 
which enters the classical gauge theory Lagrangian via the natural pairing
\begin{equation}
\begin{aligned}
{\bf I}_{\upzeta}(A) \,&=\, \frac{1}{2\pi i} \int_M \upzeta\^F_A\,,\\
&=\, \frac{1}{2\pi i} \int_M d^3x\,\sqrt{g}\,\epsilon_{m n p} \,
\upzeta^p \, F_A^{m n}\,.
\end{aligned}
\end{equation}
The partition function $Z_M$ then depends upon $\upzeta$ as a
theta-function associated to the cohomology lattice of
$M$, and abelian duality acts as a modular transformation on that
theta-function.  When ${\upzeta = 0}$, the partition function is
nonetheless invariant under duality, but in a fairly non-trivial way.

This observation appears at least implicitly in \cite{BrodaWG,ProdanovJY},
with which the present work has some overlap, but I believe it
deserves further emphasis here.  I also take the opportunity to clean
up a few factors in \cite{BrodaWG}, which otherwise detract from a
very elegant analysis.   Similar questions about one-loop determinants
and global aspects of duality have been addressed in the
supergravity literature; see for instance 
\cite{Duff:1980qv,Duff:2010ss,Grisaru:1984vk} and references therein.

\newpage
\noindent{\sl The Plan of the Paper}\medskip

Very broadly, the purpose of this paper is to analyze the quantum
analogue of the classical abelian duality relation in \eqref{DUALTD} when $M$
is a general Riemannian \mbox{three-manifold}.  Because the quantum
field theories on both sides of the duality are free, this analysis is
straightforward and can be carried out in an explicit fashion.  

In fact, I will carry out the analysis two ways, working in both the
Lagrangian and the Hamiltonian formalisms, since one learns different
things from each.  Here  I focus on the Lagrangian perspective, and in
a companion paper \cite{BeasleyII}, I adopt the alternative 
Hamiltonian viewpoint.

In Sections \ref{Scalar} and \ref{Vector}, I compute the respective
partition functions for a periodic\footnote{The adjective ``periodic''
  is traditional but possibly misleading.  More precisely, the scalar
  field will be circle-valued.} scalar field and an abelian
gauge field on the three-manifold $M$.  Then in Section \ref{Modular}, I
perform a direct comparison of the resulting expressions for the
partition function.  As mentioned above, these expressions involve a
novel theta-function attached to the three-manifold $M$, akin to the
classical theta-function on the Jacobian variety of a Riemann
surface.  Duality acts by a modular transformation on the theta-function.

Both to orient the reader and for sake of completeness, I conclude in
Section \ref{Path} by reviewing the standard path integral explanation
for abelian duality in three dimensions.  A very nice exposition of
the latter material appears in Lecture 8 of \cite{Deligne}, which I
largely follow.  I also discuss duality for three simple classes of
operators (Wilson loops, vortex loops, and monopole operators in the
language of Maxwell theory) whose commutator algebra on a Riemann
surface of genus $g$ will be analyzed in \cite{BeasleyII}.
See also \cite{Kapustin:2009av} for another recent approach to abelian duality,
invoking the formalism of duality walls.

One coupling relevant in dimension three but with no equivalent in
dimensions two and four is the Chern-Simons coupling, for which global
issues feature prominently.  In subsequent work, I apply ideas 
here and in \cite{BeasleyII} to clarify the meaning of  abelian
duality for Maxwell-Chern-Simons theory at level $k$.

\medskip
\section{Analysis of the Abelian Sigma Model}\label{Scalar}

We first compute the partition function for a free, periodic scalar
field on $M$.  Throughout this paper, $M$ is a closed, oriented
three-manifold, with Riemannian metric $g$.  The most basic topological invariant of
$M$ is the first Betti number $b_1$, which is the dimension of the
vector space $\CH^1(M)$ of harmonic one-forms on
$M$.  As for the other Betti numbers, trivially ${b_0 = b_3 = 1}$, and
  ${b_2 = b_1}$ by Poincar\'e duality.

Unlike instances of topological quantum field theory, the
abelian quantum field theories here will definitely depend upon the
choice of the Riemannian metric $g$.  The most 
elementary invariant of the metric on $M$ is the total volume,
parametrized in terms of an overall length scale $\ell$,
\begin{equation}\label{VOLM}
\ell^3 \,=\, \int_M \vol_M\,,\qquad\qquad \vol_M \,=\, \*1 \,\in\,
\Omega^3(M)\,.
\end{equation}

As we perform computations, we will wish to keep track of the
dependence on both the length scale $\ell$ and the electric coupling
$e^2$, which enters the fundamental duality relation in \eqref{DUALTD}.  This
bookkeeping is easy, for under a scale transformation 
\begin{equation}\label{SCALEg}
g \,\longmapsto\, \Lambda^2\,g\,,\qquad\qquad\qquad \Lambda
\,\in\,\BR_+\,,
\end{equation}
the parameter $\ell$ naturally transforms as 
\begin{equation}\label{SCALEl}
\ell \,\longmapsto\, \Lambda\,\ell\,.
\end{equation}
This transformation should be compared to the transformation in 
\eqref{SCALEIII} of the electric coupling,
\begin{equation}\label{SCALEe}
e^2 \,\longmapsto\, \Lambda^{-1} \, e^2\,.
\end{equation}
From \eqref{SCALEl} and \eqref{SCALEe}, we immediately see that the
dimensionless combination 
${ e^2\ell}$ is invariant under an overall rescaling of the metric on $M$.

Because the abelian quantum field theories under consideration are
free, they can always be defined so that the transformations in
\eqref{SCALEg} and \eqref{SCALEe} preserve both the classical action
and the quantum partition function on $M$.  The two parameters $e^2$ and
$\ell$ are then redundant, since either $e^2$ or $\ell$ can be scaled
to unity with an appropriate choice of ${\Lambda \in \BR_+}$.
Nevertheless, I leave the dependence on both $e^2$ and $\ell$
explicit, and invariance under scaling will be a small check on our
later formulas.

\subsection{The Classical Sigma Model}

Classically, a periodic scalar field $\phi$ on $M$ simply describes a
map from $M$ to the circle,
\begin{equation}\label{MAPPHI}
\phi: M \,\longrightarrow\, S^1\,\simeq\,\BR/2\pi\BZ\,.
\end{equation}
As indicated on the right in \eqref{MAPPHI}, we interpret $\phi$ as an angular
quantity, subject to the identification 
\begin{equation}\label{TWOPI}
\phi \,\sim\, \phi \,+\, 2\pi\,.
\end{equation}
The assumption in \eqref{MAPPHI} that $\phi$ is valued in $S^1$, as
opposed to $\BR$, has important global consequences.  

Abstractly, a given choice for $\phi$ determines a point in the space
$\CX$ of all maps from $M$ to $S^1$, 
\begin{equation}\label{BIGX}
\CX \,=\, \Map\!\left(M, S^1\right)\,.
\end{equation}
In general, $\CX$ is not connected, but rather decomposes
into components labelled by the homotopy class
of the map $\phi$. By standard facts in topology (see for instance
Chapter 4.3 in \cite{HatcherAl}), homotopy classes of maps from $M$ to
$S^1$ are in one-to-one correspondence with cohomology classes in 
$H^1(M;\BZ)$.  Under this correspondence, the cohomology class 
associated to a given map $\phi$ is the pullback to $M$ under
$\phi$ of a fixed generator for ${H^1(S^1;\BZ)\simeq\BZ}$.  Abusing 
notation slightly, I write this pullback as    
\begin{equation}\label{CLASS}
\left[\frac{d\phi}{2\pi}\right] \in H^1(M;\BZ)\,.
\end{equation}

Throughout this paper, we will treat torsion in integral cohomology
with care.  By the Universal Coefficient Theorem, the
cohomology group $H^1(M;\BZ)$ is generated freely over $\BZ$, without
torsion.  Thus $H^1(M;\BZ)$ is a lattice with rank $b_1$,
\begin{equation}\label{BIGL}
\BL \,\equiv\, H^1(M;\BZ) \,\simeq\, \BZ^{b_1}\,,
\end{equation}
where the notation $\BL$ merely serves as a convenient shorthand.  As one might
guess, the lattice $\BL$ will play a prominent role in what follows.

Altogether, the space $\CX$ in \eqref{BIGX} decomposes into connected 
components labelled by a winding-number $\omega$ which is
valued in the cohomology lattice $\BL$,
\begin{equation}\label{DECM}
\CX \,=\, \bigsqcup_{\omega\in\BL}
\CX_\omega\,.
\end{equation}
Here $\CX_\omega$ consists of those sigma model maps which satisfy 
\begin{equation}\label{WINDPHI}
\CX_\omega \,=\, \left\{\phi:M\to S^1 \,\Biggr|\, \left[\frac{d\phi}{2\pi}\right]
  \,=\,\omega\right\}\,,\qquad\omega\,\in\,\BL\,.
\end{equation}

The free sigma model action for $\phi$ takes the standard form 
\begin{equation}\label{SIGMAI}
\begin{aligned}
{\bf I}_0(\phi) \,&=\, \frac{e^2}{4\pi} \int_M d\phi \^\*d\phi\,,\\
&=\, \frac{e^2}{4\pi} \int_M \sqrt{g} \, \partial_m\phi \, \partial^m \phi \,
d^3 x\,,\qquad\qquad m\,=\,1,2,3\,.
\end{aligned}
\end{equation}
In the second line of \eqref{SIGMAI}, we write the sigma model action in local
coordinates on $M$, with the Einstein summation convention applied to
the index `$m$'.  The factor of ${1/4\pi}$ in the normalization of ${\bf I}_0$
is again a numerical convenience.

The sigma model action ${\bf I}_0$ includes a prefactor which we will 
eventually identify with the electric coupling $e^2$ under duality.
As for the discussion of the Maxwell action in Section \ref{Intro}, the
overall dependence on $e^2$ in \eqref{SIGMAI} is fixed by invariance
under the scale transformations in \eqref{SCALEg} and \eqref{SCALEe}.
Under the scaling of the metric, the field $\phi$ is necessarily invariant,
since any non-trivial scaling of $\phi$ would be incompatible with the
fixed angular identification in \eqref{TWOPI}.  Otherwise, due to its
implicit metric dependence, the dual two-form $\*d\phi$ scales as
${\*d\phi \mapsto \Lambda \, \*d\phi}$.\footnote{The scaling 
  of the two-form $\*d\phi$ is perhaps most easily examined in local
  coordinates, 
  where ${\*d\phi \equiv \sqrt{g} \, \epsilon_{m n p} \, \partial^p\phi
    \, dx^m \^ dx^n}$.  Under the scale transformation ${g \mapsto \Lambda^2
    g}$, the volume factor $\sqrt{g}$ transforms as ${\sqrt{g} \mapsto
    \Lambda^3 \sqrt{g}}$.  On the other hand, ${\partial^p \phi = g^{p
      q} \,\partial_q\phi}$ scales as ${\partial^p\phi \mapsto
    \Lambda^{-2} \, \partial^p\phi}$.  So $\*d\phi$ scales in total as 
  ${\*d\phi\mapsto \Lambda \,\*d\phi}$.}  Because the
coupling $e^2$ scales inversely to $\*d\phi$, the sigma model action
${\bf I}_0$ is thus invariant.

As usual, the metric on $M$ induces an inner-product on the space
$\Omega^p(M)$ of smooth $p$-forms for each ${p=0,\ldots,3}$,
\begin{equation}\label{LTWOP}
\left(\eta,\xi\right) \,=\, \int_M \eta\^\*
\xi\,,\qquad\qquad \eta, \xi\,\in\,\Omega^p(M)\,.
\end{equation}
In terms of the $L^2$ inner-product, the sigma model action can be abbreviated 
\begin{equation}\label{LTWOPII}
{\bf I}_0(\phi) \,=\, \frac{e^2}{4\pi} \left(d\phi, d\phi\right).
\end{equation}

More or less immediately, the critical points of the free sigma
model action in \eqref{LTWOPII} are harmonic maps from $M$ to $S^1$, 
\begin{equation}\label{ELSIG}
\delta{\bf I}_0(\phi_{\rm cl}) \,=\, 0 \qquad\Longleftrightarrow\qquad
\triangle\phi_{\rm cl} \,=\, 0\,.
\end{equation}
Because $M$ is compact, any $\BR$-valued harmonic function on $M$ is
constant and hence unique up to normalization.  For $S^1$-valued
harmonic functions as in \eqref{ELSIG}, a roughly similar statement holds.

First, by classical Hodge theory, each integral cohomology class
${\omega\in\BL}$ admits a unique harmonic representative with integral
periods on $M$.  Abusing notation slightly, I also use $\omega$ to denote
the corresponding harmonic one-form, which depends upon the Riemannian
metric on $M$.  If ${\phi_{\rm cl} \in \CX_\omega}$ is a circle-valued
harmonic map with winding-number $\omega$, then necessarily $\phi_{\rm cl}$ is
related to $\omega$ by 
\begin{equation}\label{HARMPH}
d\phi_{\rm cl} \,=\, 2\pi\omega\,,\qquad\qquad \omega\,\in\,\CH^1(M)\,.
\end{equation}
This condition implies both that $\phi_{\rm cl}$ has winding-number
$\omega$ and that $\phi_{\rm cl}$ is harmonic, since
\begin{equation}
\triangle\phi_{\rm cl} \,=\, d^\dagger d\phi_{\rm cl} \,=\, 2\pi
d^\dagger\omega \,=\, 0\,.
\end{equation}
Given the integral harmonic form $\omega$, the linear equation in
\eqref{HARMPH} can always be solved and so determines $\phi_{\rm cl}$
up to the addition of a constant.  As a result, the moduli space of
harmonic maps with winding-number $\omega$ is a copy of $S^1$.

\medskip\noindent{\sl A Topological Parameter for the Sigma
  Model}\smallskip

Given the decomposition for $\CX$ in \eqref{DECM}, we naturally
extend the free sigma model action ${\bf I}_0(\phi)$ to include a
topological term which is locally-constant on $\CX$ and hence only
sensitive to the winding-number $\omega$.  

The most obvious topological term for $\phi$ depends upon the choice of a
de Rham cohomology class 
\begin{equation}\label{TOPPAR}
\upbeta \,\in\, H^2(M;\BR) \,\simeq\,\BR^{b_1}\,,
\end{equation}
with the pairing 
\begin{equation}\label{TOPPHI}
{\bf I}_\upbeta(\phi) \,=\, \frac{1}{2\pi i}\int_M \upbeta \^
d\phi\,.
\end{equation}
Because $\upbeta$ and $d\phi$ are both closed forms on the compact 
manifold $M$, the pairing in \eqref{TOPPHI} depends only on the
cohomology class of $\upbeta$ and on the homotopy class of the map
$\phi$.  In particular, ${\bf I}_\upbeta(\phi)$ is unchanged under any
variation of $\phi$, so the addition of ${\bf I}_\upbeta(\phi)$ to the
sigma model action does not change the harmonic equation of motion for
$\phi$.

One might suppose that ${\bf I}_\upbeta(\phi)$ is the end of the
story, since there are hardly any other topological couplings to write
for the free sigma model.  However, this paper is about duality, and
the choice of $\upbeta$ in \eqref{TOPPAR} does not respect the 
fundamental duality on $M$ --- to wit, Poincar\'e duality.  

The most elegant formulation of the scalar partition function on $M$
occurs when we introduce a second cohomological parameter dual to
$\upbeta$.  The dual parameter $\upalpha$ is a harmonic one-form on
$M$,
\begin{equation}\label{TOPAP}
\upalpha \,\in\, \CH^1(M) \,\simeq\, \BR^{b_1}\,,
\end{equation}
which couples linearly to the sigma model map $\phi$ via 
\begin{equation}\label{IALPHA}
{\bf I}_\upalpha(\phi) \,=\, \frac{e^2}{2\pi} \int_M \*\upalpha\^d\phi\,.
\end{equation}
The choice of Riemannian metric on $M$ enters both the definition of
$\upalpha$ as a harmonic one-form and the description of the 
coupling ${\bf I}_\upalpha(\phi)$ in \eqref{IALPHA}.  However, precisely because
$\*\upalpha$ is closed, the value of ${\bf I}_\upalpha(\phi)$ does not
change under variations of $\phi$, so ${\bf I}_\upalpha(\phi)$ is also
a locally-constant function on the space $\CX$ of sigma model maps.
Hence ${\bf I}_\upalpha(\phi)$ 
does not alter the harmonic equation of motion for $\phi$ either.  We
include the prefactor of $e^2$ in \eqref{IALPHA} to ensure that ${\bf
  I}_\upalpha(\phi)$ is invariant under the scale transformations in
\eqref{SCALEg} and \eqref{SCALEe}.

Including both the cohomological parameters $\upalpha$ and $\upbeta$,
the total sigma model action becomes 
\begin{equation}\label{ITOTI}
\begin{aligned}
{\bf I}_{\rm tot}(\phi) \,&=\, {\bf I}_0(\phi) \,+\, {\bf
  I}_\upbeta(\phi) \,+\, {\bf 
  I}_\upalpha(\phi)\,,\\
&=\,  {\bf I}_0(\phi) \,+\, \frac{1}{2\pi i} \int_M\!\left(\upbeta +
  i\,e^2\*\upalpha\right)\!\^d\phi\,.
\end{aligned}
\end{equation}
Though $\upalpha$ and $\upbeta$ enter the total action 
similarly, an asymmetry exists in our description of these parameters.  The
parameter $\upalpha$ is a definite harmonic one-form 
on $M$, whereas $\upbeta$ is any closed two-form representing the given
cohomology class.

A loose theme running throughout this work and its companion
\cite{BeasleyII} is the interpretation of abelian duality on $M$ as a 
kind of quantum Poincar\'e duality, which here will exchange
$\upalpha$ and $\upbeta$.  To make sense of this exchange, though, we
will need to select a definite two-form on $M$ to represent the
cohomology class of $\upbeta$.  Because ${\upalpha\in\CH^1(M)}$ is
already harmonic, we also use the metric on $M$ to determine a harmonic
representative ${\upbeta \in \CH^2(M)}$ for the cohomology class in
\eqref{TOPPAR}.

With this choice, the sigma model action in \eqref{ITOTI} only depends
on the parameters $\upalpha$ and $\upbeta$ in the holomorphic combination
\begin{equation}\label{BZETA}
\upgamma \,=\, \upbeta \,+\, i \, e^2\,\*\upalpha \,\in\, 
\CH^2_\BC(M) \,\simeq\, \BC^{b_1}\,,
\end{equation}
a complex harmonic two-form on $M$.  By construction, $\upgamma$ is
invariant under the combined scalings of the metric $g$ and coupling
$e^2$ in \eqref{SCALEg} and \eqref{SCALEe}.

The total sigma model action can then be written concisely as 
\begin{equation}\label{TOTAC}
{\bf I}_{\rm tot}(\phi) \,=\, \frac{e^2}{4\pi}\left(d\phi,d\phi\right)
\,+\, \frac{1}{2\pi i} \left\langle\upgamma,d\phi\right\rangle,
\end{equation}
where $\left(\,\cdot\,,\,\cdot\,\right)$ is the $L^2$ inner-product, and
$\left\langle\,\cdot\,,\,\cdot\,\right\rangle$ is the canonical
intersection pairing,
\begin{equation}\label{DUALP}
\left\langle\eta,\xi\right\rangle \,=\, \int_M
\eta\^\xi\,,\qquad\qquad \eta\,,\xi\,\in\,\Omega^*(M)\,.
\end{equation}

\subsection{Computing the Partition Function}\label{ScalarP}

We now evaluate the partition function for the periodic
scalar field on $M$ using the path integral presentation 
\begin{equation}\label{ZPHI}
Z_M(\upgamma) \,=\,
\sum_{\omega\in\BL}\,\int_{\CX_\omega}\!\!\CD\phi\,\exp{\!\big[-{\bf
    I}_{\rm tot}(\phi)\big]}\,.
\end{equation}
Here I indicate the explicit dependence of the partition function
on the holomorphic parameter ${\upgamma \in \CH^2_\BC(M)}$, and
I leave implicit the combined dependence on the coupling $e^2$ and the
Riemannian metric $g$.

Because the sigma model configuration space $\CX$ decomposes into
components labelled by the winding-number
${\omega\in\BL}$, the sigma model path integral includes a sum over
the cohomology lattice $\BL$, followed by an integral 
over each component $\CX_\omega \subset \CX$.  With the free sigma
model action in \eqref{TOTAC}, those integrals are all Gaussian and hold
no particular mysteries.  However, we would like to assign a definite
normalization to $Z_M(\upgamma)$, and for that goal, we must be careful about
the normalization of the sigma model measure $\CD\phi$ itself.

\medskip\noindent{\sl More about the Sigma Model Measure}\medskip

Although the configuration space ${\CX = \Map(M,S^1)}$ decomposes into
an infinite number of components $\CX_\omega$ labelled by the winding-number
$\omega$, each component can be identified
with the distinguished component $\CX_0$, which consists of 
maps with trivial winding.  To make the
identification ${\CX_\omega \simeq \CX_0}$, we select a basepoint
${\Phi_\omega \in \CX_\omega}$, corresponding to a fiducial map with
winding-number $\omega$.  Given the fiducial map ${\Phi_\omega}$, any
other map ${\phi\in\CX_\omega}$ with the same winding-number can be written
uniquely as a sum 
\begin{equation}\label{FIDUC}
\phi \,=\, \Phi_\omega \,+\, \psi\,,\qquad\qquad \psi \,\in\,\CX_0\,,
\end{equation}
where ${\psi:M\to S^1}$ is a sigma model map with vanishing winding.  The correspondence
between $\phi$ and $\psi$ in \eqref{FIDUC} provides the desired
identification of $\CX_\omega$ and $\CX_0$.  As an immediate corollary, if we wish to characterize
the sigma model measure $\CD\phi$ on $\CX_\omega$, we need only
characterize it on $\CX_0$.

We will characterize the measure on $\CX_0$ momentarily, but let us
first make a definite choice for the fiducial map ${\Phi_\omega
  \in \CX_\omega}$ in \eqref{FIDUC}.  Our choice will
depend upon the Riemannian metric $g$ on $M$, as well as the data of
a point ${p\in M}$.  Using the metric, we first impose the condition
that ${\Phi_\omega:M\to S^1}$ be harmonic, or equivalently
\begin{equation}\label{FIDMAP}
d\Phi_\omega \,=\, 2 \pi \omega\,,\qquad\qquad \omega\,\in\,\CH^1(M)\,.
\end{equation}
The condition in \eqref{FIDMAP} determines $\Phi_\omega$ up to the
addition of a constant.  To fix the constant, we next impose 
\begin{equation}\label{BASEP}
\Phi_\omega(p) \,=\, 0 \,\,\mod\,\, 2\pi\,,\qquad\qquad p\,\in\, M\,.
\end{equation}
Together, the conditions in \eqref{FIDMAP} and \eqref{BASEP} determine
the map $\Phi_\omega$ uniquely.

At any given point ${\phi \in \CX_0}$, the tangent space to $\CX_0$ at
$\phi$ is simply the space of real-valued functions $\Omega^0(M)$,
\begin{equation}
T_{[\phi]}\CX_0 \,=\, \Omega^0(M)\,.
\end{equation}
Indeed globally,
\begin{equation}
\CX_0 \,=\, \Omega^0(M) \,\,\mod\,\, 2\pi\,.
\end{equation} 
The metric on $M$ immediately induces a metric on $\CX_0$, given by
\begin{equation}\label{BIGM}
||\delta\phi||^2_{\CX_0} \,=\, \frac{e^6}{\left(2\pi\right)^2} \int_M
\delta\phi\^\*\delta\phi\,,\qquad\qquad \delta\phi\,\in\,\Omega^0(M)\,.
\end{equation}
The appearance of the $L^2$-norm on $\Omega^0(M)$ should come as no
surprise, but the coupling-dependent prefactor in \eqref{BIGM} may
be one.  Under a scaling of the metric ${g \mapsto \Lambda^2\, g}$, the standard
$L^2$-norm on $\Omega^0(M)$ scales according to the volume of $M$, 
\begin{equation}
\int_M \delta\phi\^\*\delta\phi \,\longmapsto\, \Lambda^3 \int_M
\delta\phi\^\*\delta\phi\,,\qquad\qquad \Lambda\,\in\,\BR_+\,.
\end{equation}
Since ${e^2 \mapsto \Lambda^{-1} \, e^2}$ according to \eqref{SCALEe}, the
prefactor of $e^6$ in $||\delta\phi||^2_{\CX_0}$ ensures invariance of the metric on
$\CX_0$ under scaling.  The remaining factors of $2\pi$ in
\eqref{BIGM} appear by convention.

Once $\CX_0$ carries a Riemannian structure, we take $\CD\phi$ to be
the corresponding Riemannian measure.  Under the identification
${\CX_\omega \simeq \CX_0}$, the sigma model measure 
then extends to all of the configuration space $\CX$.  Finally, by
construction $\CD\phi$ is invariant under translations by elements in
$\Omega^0(M)$.  Hence 
$\CD\phi$ does not actually depend upon the particular choice of
basepoint ${\Phi_\omega\in\CX_\omega}$ specified in \eqref{FIDMAP} and
\eqref{BASEP}.  

Though the description of $\CD\phi$ in terms of the
metric on $\CX_0$ is purely formal, we will see later that this
description, formal or no, allows us to perform a precise accounting
under duality of all coupling-dependent factors in the partition
function.  This accounting clarifies the results in \cite{BrodaWG}.

\medskip\noindent{\sl A Sum over Windings}\medskip

To evaluate the partition function on $M$,
\begin{equation}\label{ZPHTWO}
Z_M(\upgamma) \,=\,
\sum_{\omega\in\BL}\,\int_{\CX_\omega}\!\!\CD\phi\,\exp{\!\big[-{\bf
    I}_{\rm tot}(\phi)\big]}\,,
\end{equation}
we first employ the componentwise identification ${\CX_\omega \simeq
  \CX_0}$ to rewrite the integration variable $\phi$ in \eqref{ZPHTWO}
as the sum 
\begin{equation}\label{PHISI}
\phi \,=\, \Phi_\omega \,+\, \psi\,,\qquad\qquad \Phi_\omega \,\in\,
\CX_\omega\,,\qquad\qquad \psi \in \CX_0\,.
\end{equation}
With the substitution in \eqref{PHISI}, the partition function can be
computed in terms of a path integral over the distinguished component
$\CX_0$ alone,
\begin{equation}\label{SUMZPHII}
Z_M(\upgamma) \,=\, \sum_{\omega\in\BL}
\,\int_{\CX_0}\!\!\CD\psi\,\exp{\!\big[-{\bf I}_{\rm
    tot}\!\left(\Phi_\omega + \psi\right)\big]}\,.
\end{equation}
In general, when making changes of variables in the path integral, one
must be careful about Jacobians, but according to our preceding
description of $\CD\phi$, the Jacobian for the
substitution in \eqref{SUMZPHII} is unity.

In terms of the harmonic map $\Phi_\omega$ and the
homotopically-trivial map $\psi$, the sigma model action becomes 
\begin{equation}\label{ITOTIV}
{\bf I}_{\rm tot}\!\left(\Phi_\omega + \psi\right)
\,=\, \frac{e^2}{4\pi}\!\left(d\Phi_\omega + d\psi, d\Phi_\omega + d\psi\right)
\,+\, \frac{1}{2\pi i}\left\langle\upgamma, d\Phi_\omega +
d\psi\right\rangle\,,
\end{equation}
or more explicitly,
\begin{equation}\label{ITOTV}
{\bf I}_{\rm tot}\!\left(\Phi_\omega + \psi\right)
\,=\,\frac{e^2}{4\pi}\int_M \left(2\pi\omega \,+\, 
d\psi\big)\!\^\*\!\big(2\pi\omega \,+\, d\psi\right) + \,\frac{1}{2\pi
  i} \int_M \upgamma \^\!\left(2\pi\omega \,+\, d\psi\right)\,.
\end{equation}
In passing to \eqref{ITOTV}, I recall that the fiducial map satisfies
${d\Phi_\omega = 2\pi\omega}$.  Since both $\omega$ and $\upgamma$ are harmonic
forms on $M$, the cross-terms in \eqref{ITOTV} which involve
either $\omega$ or $\upgamma$ together with $d\psi$ vanish, so that 
\begin{equation}\label{SUMIT}
\begin{aligned}
{\bf I}_{\rm tot}\!\left(\Phi_\omega + \psi\right) 
\,&=\, \frac{e^2}{4\pi} \int_M d\psi \^\*d\psi  \,+\, \pi e^2 
\int_M \omega\^\*\omega \,-\, i \int_M \omega\^\upgamma\,,\\
&=\, \frac{e^2}{4\pi}\left(d\psi,d\psi\right) + \pi e^2
\left(\omega,\omega\right) \,-\, i \left\langle\omega,\upgamma\right\rangle.
\end{aligned}
\end{equation}

From the description of the sigma model action in \eqref{SUMIT}, the
partition function on $M$ can be rewritten more explicitly as 
\begin{equation}\label{SUMMERZ}
Z_M(\upgamma)
\,=\,\sum_{\omega\in\BL}\int_{\CX_0}\!\!\CD\psi\,
\exp{\!\left[-\frac{e^2}{4\pi}\left(d\psi,d\psi\right) - \pi e^2
    \left(\omega,\,\omega\right) \,+\, i
    \left\langle\omega,\upgamma\right\rangle\right]}\,. 
\end{equation}
Because the argument of the exponential in
\eqref{SUMMERZ} is a sum of terms which depend separately 
on the variables ${\omega\in\BL}$ and ${\psi\in\CX_0}$, the partition function
immediately factorizes,
\begin{equation}\label{FACTZM}
Z_M(\upgamma) \,=\, \Delta_M \cdot \Theta_M(\upgamma)\,,
\end{equation}
where $\Delta_M$ is given by a Gaussian integral over the space
$\CX_0$ of homotopically-trivial maps ${\psi:M\to S^1}$,
\begin{equation}
\Delta_M \,=\,
\int_{\CX_0}\!\!\CD\psi\,
\exp{\!\left[-\frac{e^2}{4\pi}\left(d\psi,d\psi\right)\right]}\,,
\end{equation}
and $\Theta_M(\upgamma)$ is given by a discrete sum over the
cohomology lattice ${\BL = H^1(M;\BZ)}$,
\begin{equation}\label{BIGXI}
\Theta_M(\upgamma) \,=\, \sum_{\omega\in\BL}\,\exp{\!\left[-\pi
    e^2 \left(\omega,\,\omega\right) \,+\, i
    \left\langle\omega,\upgamma\right\rangle\right]}\,.
\end{equation}
Due to the Gaussian damping, the lattice sum which defines
$\Theta_M(\upgamma)$ in \eqref{BIGXI} is convergent for all
${\upgamma \in \CH^2_\BC(M)}$.

From the perspective of duality, the more interesting term in the
factorization \eqref{FACTZM} is $\Theta_M(\upgamma)$, which carries the
dependence on the holomorphic parameter $\upgamma$ and arises from the
quantum sum over winding-sectors in the sigma model.  Clearly 
$\Theta_M(\upgamma)$ is a theta-function attached to the
three-manifold $M$ though the cohomology lattice ${\BL =
  H^1(M;\BZ)}$, and wherever a theta-function appears, the modular
group lurks.  For the time being, though, I postpone discussion of
$\Theta_M(\upgamma)$ and its role in duality until Section \ref{Modular}.

\medskip\noindent{\sl The Normalization of $Z_M(\upgamma)$}\medskip

Finally, to fix the absolute normalization of the scalar partition
function, we are left to compute the factor 
\begin{equation}\label{DELMII}
\Delta_M \,=\,
\int_{\CX_0}\!\!\CD\psi\,
\exp{\!\left[-\frac{e^2}{4\pi}\,\big(d\psi,d\psi\big)\right]}\,,
\end{equation}
depending on the coupling $e^2$ and the Riemannian metric on $M$.
Because the path integral in \eqref{DELMII} is Gaussian, the only
trick will be to keep track of factors associated to the normalization
of the path integral measure $\CD\psi$.

As a first step, we expand $\psi$ in an orthonormal basis for ${\CX_0 \simeq
  \Omega^0(M)}$ mod $2\pi$ with respect to the metric in \eqref{BIGM},
\begin{equation}\label{SPECTR}
\psi \,=\, \psi_0 \cdot
\left[\frac{2\pi}{\left(e^2\ell\right)^{3/2}}\right] + \sum_{\lambda >
  0} \, \psi_\lambda \cdot \hat{f}_\lambda\,.
\end{equation}
Here each $\hat{f}_\lambda$ is a normalized eigenfunction with
eigenvalue $\lambda$ for the scalar Laplacian $\triangle$ on $M$,
\begin{equation}\label{EFUN}
\triangle \hat{f}_\lambda \,=\, \lambda \, \hat{f}_\lambda\,,\qquad\qquad
||\hat{f}_\lambda||^2_{\CX_0} \,=\, \frac{e^6}{\left(2\pi\right)^2}\int_M
\hat{f}_\lambda^2 \,\, \vol_M \,=\, 1\,,
\end{equation}
and each spectral coefficient $\psi_\lambda$ is valued in $\BR$.  

More important is the coefficient of the constant function 
\begin{equation}
\hat{f}_0 \,=\,
\left[\frac{2\pi}{\left(e^2\ell\right)^{3/2}}\right]\,,\qquad\qquad
||\hat{f}_0||^2_{\CX_0} \,=\, \frac{e^6}{\left(2\pi\right)^2}\int_M
\hat{f}_0^2 \,\, \vol_M \,=\, 1\,,
\end{equation}
also chosen to have unit-norm in $\Omega^0(M)$.  Because $\psi$ 
satisfies the periodicity condition ${\psi \sim \psi + 2\pi}$, the
coefficient $\psi_0$ of the constant mode $\hat{f}_0$ must
have period 
\begin{equation}\label{ZMPER}
\psi_0 \,\sim\, \psi_0 \,+\, (e^2\ell){}^{3/2}\,.
\end{equation}
Though we have fixed the periodicity of the sigma model
field $\psi$ to be independent of the coupling, the effective
periodicity of the zero-mode $\psi_0$ does depend upon $e^2$ and diverges as
${e^2\to\infty}$ with the volume of $M$ held fixed.

In terms of the spectral decomposition \eqref{SPECTR} for $\psi$,
the Gaussian path integral in \eqref{DELMII} becomes 
\begin{equation}\label{DETM}
\Delta_M \,=\,\int_{\CX_0}\!\!d\psi_0\,\CD\psi'\,
\exp{\!\left[-\frac{e^2}{4\pi}\,\big(\psi',\triangle\psi'\big)\right]}\,.
\end{equation}
Here $\psi'$ indicates the orthocomplement to the constant mode in
$\Omega^0(M)$, 
\begin{equation}\label{PSIPR}
\psi' \,=\, \sum_{\lambda > 0} \, \psi_\lambda \cdot
\hat{f}_\lambda\,,\qquad\qquad \psi_\lambda\,\in\,\BR\,,
\end{equation}
and at least formally, $\CD\psi'$ is shorthand for the product measure 
\begin{equation}
\CD\psi' \,=\, \prod_{\lambda > 0} \,d\psi_\lambda\,.
\end{equation}
As standard, in obtaining \eqref{DETM} we integrate by parts to
produce the scalar Laplacian ${\triangle = 
  d^\dagger d}$ acting on $\psi'$ in the argument of the exponential.

Again with care for factors of $e^2$, we apply the expansion of
$\psi'$ in \eqref{PSIPR} to simplify that argument,
\begin{equation}\label{PSIPRII}
\begin{aligned}
\frac{e^2}{4\pi}\left(\psi',\triangle\psi'\right) \,&=\,
\frac{e^2}{4\pi}\,\sum_{\lambda > 0} \, \lambda \, \psi_\lambda^2 \,
(\hat{f}_\lambda, \hat{f}_\lambda)\,,\\
&=\, \frac{\pi}{e^4} \, \sum_{\lambda > 0} \, \lambda \, \psi_\lambda^2\,.
\end{aligned}
\end{equation}
In passing to the second line of \eqref{PSIPRII}, we
observe that the eigenfunction $\hat{f}_\lambda$ has $L^2$-norm
${(\hat{f}_\lambda,\hat{f}_\lambda) = 4\pi^2/e^6}$ according to
\eqref{EFUN}.  

So altogether,
\begin{equation}\label{ONDME}
\Delta_M \,=\, \int_{\CX_0}
\!\!d\psi_0\,\CD\psi'\,\exp{\!\left[-\frac{\pi}{e^4}\sum_{\lambda > 0}
  \lambda \, \psi_\lambda^2\right]}\,,
\end{equation}
or upon evaluating the Gaussian integrals over each spectral
coefficient $\psi_\lambda$,
\begin{equation}\label{DMFIN}
\Delta_M \,=\, (e^2\ell){}^{3/2} \cdot
\frac{1}{\sqrt{\det'\!\left(\triangle/e^4\right)}}\,.
\end{equation}
The slightly tricky prefactor $\left(e^2\ell\right)^{3/2}$ arises from
the integral over the zero-mode $\psi_0$ with the periodicity in
\eqref{ZMPER}, and $\det'$ indicates the determinant with kernel
omitted.  Some regularization method must be chosen to define the 
functional determinant in \eqref{DMFIN} as a real number, with
zeta-function regularization being one possibility.   See for instance
\cite{Friedmann:2002ty,Nash:1992sf} for explicit calculations of such
zeta-regularized determinants on lens spaces ${M = S^3/\BZ_k}$, with
the round metric inherited from $S^3$.

The expression for $\Delta_M$ in \eqref{DMFIN} makes manifest the fact
that $\Delta_M$ is invariant under the simultaneous scalings of the
metric and coupling in \eqref{SCALEg} and \eqref{SCALEe}.  In
particular, one can easily check that the operator $\triangle/e^4$ is
invariant under scaling, since $\triangle$ scales with $\Lambda$ as
${\triangle \mapsto \Lambda^{-2}\,\triangle}$.  However, the appearance of
the coupling in the functional determinant $\det'(\triangle/e^4)$ is
slightly awkward, and if one wishes, the dependence on $e^2$ in
$\Delta_M$ can be made more explicit by pulling $e^2$ out from the determinant.
The same issue arises in Appendix A of \cite{Friedmann:2002ty}, whose
strategy of analysis we follow.

In zeta-regularization, the functional determinant is defined in terms
of the zeta-function for the scalar Laplacian $\triangle$ on $M$,
\begin{equation}\label{ZETAM}
\zeta_\triangle(s) \,=\, \sum_{\lambda>0} \,
\lambda^{-s}\,,\qquad\qquad s\,\in\,\BC\,. 
\end{equation}
The sum over positive eigenvalues in \eqref{ZETAM} is convergent when
the real part of $s$ is sufficiently large, and $\zeta_\triangle(s)$ is
defined for other values of $s$ by analytic continuation.  By standard
manipulations, the functional determinant of $\triangle$ is defined
in terms of the derivative of $\zeta_\triangle(s)$ at ${s=0}$,
\begin{equation}
{\det}'(\triangle) \,=\, \exp{\!\left[-\zeta'_\triangle(0)\right]}\,.
\end{equation}

Because we are interested in the determinant of the operator
$\triangle/e^4$, we instead consider 
\begin{equation}\label{ETAM}
\eta_\triangle(s) \,=\, \sum_{\lambda>0}
\left(\frac{\lambda}{e^4}\right)^{-s} \,=\, e^{4 s} \cdot \zeta_\triangle(s)\,.
\end{equation}
Then similarly, 
\begin{equation}
{\det}'(\triangle/e^4) \,=\, \exp{\!\left[-\eta'_\triangle(0)\right]}\,.
\end{equation}
On the other hand, directly from \eqref{ETAM},
\begin{equation}
\eta'_\triangle(0) \,=\, \ln(e^4) \cdot \zeta_\triangle(0) \,+\,
\zeta'_\triangle(0)\,,
\end{equation}
from which we obtain the relation 
\begin{equation}\label{SCALED}
{\det}'(\triangle/e^4) \,=\, e^{-4\,\zeta_\triangle(0)} \cdot
{\det}'(\triangle)\,.
\end{equation}

The value of $\zeta_\triangle(s)$ at ${s=0}$ can be interpreted as a
regularized dimension for the non-zero eigenspace of the operator
$\triangle$.  Very generally, if $M$ is any compact manifold of odd dimension,
and $\triangle_p$ is the de Rham Laplacian acting on the space
$\Omega^p(M)$ of $p$-forms, then the value of the associated zeta-function at
${s=0}$ is given by 
\begin{equation}\label{ZETAZ}
\zeta_{\triangle_p}(0) \,=\, -\dim\Ker\,\triangle_p\,.
\end{equation}
See Theorem $5.2$ in \cite{RosenbergS:98k} for a textbook proof of
\eqref{ZETAZ}, which goes back to \cite{Mina:49}.  For us, the scalar Laplacian
on $M$ has a one-dimensional kernel, so ${\zeta_\triangle(0)=-1}$ in
\eqref{SCALED}.  We thence obtain 
\begin{equation}
{\det}'(\triangle/e^4) \,=\, e^4 \cdot
{\det}'(\triangle)\,.
\end{equation}  
As a result, the normalization factor in \eqref{DMFIN} reduces to 
\begin{equation}\label{DMFINII}
\Delta_M \,=\, \frac{e\,\ell^{3/2}}{\sqrt{\det'\!\left(\triangle\right)}}\,.
\end{equation}
The linear dependence of $\Delta_M$ on $e$ also follows by well-known
physical arguments involving the counting of zero-modes for the
Laplacian.

\medskip\noindent{\sl The Final Result}\medskip

In summary, we have determined the scalar partition function on $M$ to be 
\begin{equation}\label{FINSCZ}
\begin{aligned}
Z_M(\upgamma) \,&=\, \Delta_M \cdot
\Theta_M(\upgamma)\,,\qquad\qquad\qquad \upgamma \,\in\,
\CH^2_\BC(M)\,\simeq\,\BC^{b_1}\,,\\ 
&=\, \frac{e\,\ell^{3/2}}{\sqrt{\det'\!\left(\triangle_0\right)}}
\cdot  \sum_{\omega\in\BL}\,\exp{\!\left[-\pi
    e^2 \left(\omega,\,\omega\right) \,+\, i
    \left\langle\omega,\upgamma\right\rangle\right]}\,.
\end{aligned}
\end{equation}
 The subscript serves to emphasize that $\triangle_0$
is the scalar Laplacian, acting on forms of degree zero.  We will
similarly meet the de Rham Laplacian $\triangle_1$ for one-forms when
we consider abelian gauge theory in Section \ref{Vector}.

\medskip
\section{Analysis of the Abelian Gauge Theory}\label{Vector}

Just as we computed the partition function for a periodic scalar
field, we now compute the partition function for an abelian gauge
field on the closed three-manifold $M$.

\subsection{The Classical Maxwell Theory}

Classically, the gauge field $A$ is a connection on a fixed principal
$U(1)$-bundle $P$ over $M$,
\begin{equation}
\begin{aligned}
U(1) \rightarrow\,\,\, &P\\[-8 pt]
&\mskip -5mu \downarrow\\[-1 ex]
&\mskip -6mu M
\end{aligned}\quad.
\end{equation}
The typical three-manifold admits many choices for the bundle $P$,
whose topological type is characterized by the first Chern class
\begin{equation}\label{CONE}
c_1\!\left(P\right) \,\in\, H^2(M;\BZ)\,.
\end{equation}
Via \eqref{CONE}, the possible types of $U(1)$-bundles on $M$ are in
one-to-one correspondence with elements of $H^2(M;\BZ)$.

In general, the abelian group $H^2(M;\BZ)$ contains torsion, and
$c_1(P)$ may be a torsion class, of finite-order in $H^2(M;\BZ)$.
Throughout this work, we will be fastidious about torsion, so we
recall the exact sequence 
\begin{equation}\label{H2SEQ}
0 \,\longrightarrow\, H^2(M;\BZ)_{\rm tors} \,\longrightarrow\,
H^2(M;\BZ) \,\longrightarrow\, H^2(M;\BZ)_{\rm free}
\,\longrightarrow\, 0\,.
\end{equation}
Here $H^2(M;\BZ)_{\rm tors}$ is the subgroup of torsion classes, and
$H^2(M;\BZ)_{\rm free}$ is the reduction of $H^2(M;\BZ)$ modulo
torsion.  Poincar\'e duality asserts that the lattice
${\BL=H^1(M;\BZ)}$ is dual under the
intersection pairing to the quotient lattice 
\begin{equation}
\BL^\vee \,=\, H^2(M;\BZ)_{\rm free} \,\simeq\, \BZ^{b_1}\,.
\end{equation}
The lattice $\BL^\vee$ also embeds in the vector space $\CH^2(M)$
as the set of harmonic two-forms with integral periods on $M$,
\begin{equation}\label{EMBLVE}
\BL^\vee \,\subset\, \CH^2(M) \,\simeq\, \BR^{b_1}\,.
\end{equation}
We will often assume implicitly the embedding in \eqref{EMBLVE}.

Though the exact sequence in \eqref{H2SEQ} can be split, the sequence does
not split in any natural way, so we should not think about $\BL^\vee$ as a
subgroup of $H^2(M;\BZ)$.  However, we can always consider the
reduction of any class in $H^2(M;\BZ)$ modulo torsion, to obtain a
class valued in $\BL^\vee$.  Throughout this paper, we normalize the gauge  
field $A$ so that the reduction of ${c_1(P)\in H^2(M;\BZ)}$ modulo torsion
admits the de Rham representative 
\begin{equation}\label{CHRNC}
\left[\frac{F_A}{2\pi}\right] \,\in\, \BL^\vee = H^2(M;\BZ)_{\rm free}\,.
\end{equation}
As usual, ${F_A = dA}$ is the curvature of the connection.  

With the normalization in \eqref{CHRNC}, homotopically non-trivial
gauge transformations act on $A$ by shifts 
\begin{equation}\label{BIGG}
A \,\longmapsto\, A \,+\, 2\pi\omega\,,\qquad\qquad \omega\,\in\,\BL =
H^1(M;\BZ)\,.
\end{equation}
Such shifts preserve the holonomy of $A$, as measured physically by
the Wilson loop operator $\SW_n(C)$ attached to an oriented closed curve ${C
  \subset M}$,
\begin{equation}\label{WILSON}
\SW_n(C) \,=\, \exp{\!\left[i\,n\oint_C A\right]}\,,\qquad\qquad
n\,\in\,\BZ\,.
\end{equation}

We have already seen that the lattice $\BL$ plays an
important role in characterizing the winding-number of the 
circle-valued map ${\phi:M\to S^1}$.  The dual lattice $\BL^\vee$
plays a similar role for Maxwell theory, since $\BL^\vee$ 
determines the topology of the $U(1)$-bundle $P$, at least up to
torsion.  Given the canonical pairing between $\BL$ and $\BL^\vee$, one
might wonder whether it is even necessary in the context of abelian
duality to consider bundles for which $c_1(P)$ is torsion.
As we will see in Section \ref{Modular}, following the original 
observation in \cite{BrodaWG}, a precise understanding of 
duality indeed requires that we consider all possibilities for
$c_1(P)$ in $H^2(M;\BZ)$, including torsion classes.

Once the bundle $P$ is fixed, we introduce the Maxwell action
 \begin{equation}\label{SMAXM}
\begin{aligned}
{\bf I}_0(A) \,&=\, \frac{1}{4\pi e^2} \int_M F_A
\^\* F_A\,,\cr
&=\, \frac{1}{8\pi e^2} \int_M \sqrt{g} \,\, F_{A, m
n}^{} \, F_A^{\,\,\,m n}\,d^3x\,,\qquad\qquad m,n=1,2,3\,.
\end{aligned}
\end{equation}
Here $e^2$ is the electric coupling, and the factor of $1/4\pi$ is
required to match our previous conventions for the scalar field.  In
the second line of \eqref{SMAXM}, we recall the component expansion of
the Maxwell action in local coordinates on $M$.

Critical points of ${\bf I}_0(A)$ correspond to connections $A_{\rm
  cl}$ on $P$ whose curvatures satisfy 
\begin{equation}\label{MAXWEQ}
\delta{\bf I}_0(A_{\rm cl}) \,=\, 0 \qquad\Longleftrightarrow\qquad
d^\dagger F_{A_{\rm cl}} \,=\, 0\,.
\end{equation}
Trivially ${d F_{A_{\rm cl}} = 0}$, so any connection which solves
the Maxwell equation on $M$ has harmonic curvature 
\begin{equation}
F_{A_{\rm cl}} \,\in\, \CH^2(M)\,.
\end{equation}
The integrality condition in \eqref{CHRNC} then determines $F_{A_{\rm
    cl}}$ uniquely according to the topological type of $P$.  For
future reference, we set 
\begin{equation}\label{HARMFA}
F_{A_{\rm cl}} \,=\, 2\pi\lambda\,,\qquad\qquad \lambda\,\in\,\BL^\vee =
H^2(M;\BZ)_{\rm free}\,.
\end{equation}

Although flux quantization restricts the classical values for $F_A$ to
a discrete set, the moduli space of classical solutions to the Maxwell
equation on $M$ generally has positive dimension whenever ${b_1 >
  0}$.  Clearly, if $A$ solves the Maxwell equation with
harmonic curvature and ${\eta\in\Omega^1(M)}$ is any closed
one-form, then ${A + \eta}$ also solves the Maxwell equation
with the same curvature.  Modulo gauge equivalence, the closed
one-form $\eta$ determines a point in the real torus 
\begin{equation}\label{JACM}
\SJ_M \,=\, H^1(M;\BR)/2\pi\BL\,\simeq\, U(1)^{b_1}\,,
\end{equation}
where we have been careful to divide by the large gauge equivalences
in \eqref{BIGG}.

For each $U(1)$-bundle $P$ over $M$, the moduli space of classical
solutions to the Maxwell equation is simply a copy of the torus
$\SJ_M$ in \eqref{JACM}.  As mentioned in the Introduction, this
moduli space bears no resemblance to the moduli space of harmonic
maps ${\phi:M\to S^1}$, which is instead a copy of $S^1$ for each
winding-sector.  Abelian duality on $M$ must therefore
involve a non-trivial quantum equivalence, even though the field
theories involved are free.  I will develop this theme further 
in \cite{BeasleyII}, where I discuss the canonical quantization of
these theories.

\medskip\noindent{\sl Adding Topological Couplings}\medskip

Just as for the periodic scalar field in Section \ref{Scalar}, we now
extend the classical Maxwell action to include additional couplings 
which will be topological in the sense that they do not change the
classical Maxwell equation in \eqref{MAXWEQ}.

The more obvious topological coupling depends upon the choice of a de
Rham cohomology class 
\begin{equation}
\upalpha \,\in\, H^1(M;\BR) \,\simeq\, \BR^{b_1}\,,
\end{equation}
for which we introduce the pairing 
\begin{equation}\label{IUPAAA}
{\bf I}_\upalpha(A) \,=\, -\frac{1}{2\pi i}\,\int_M \upalpha \^
F_A\,.
\end{equation}
Because both $\upalpha$ and $F_A$ are closed forms, the pairing in
\eqref{IUPAAA} depends only on the respective cohomology classes of
those forms.  In particular, via the identification \eqref{CHRNC},
the value of ${\bf I}_\upalpha(A)$ depends only on the image
of the Chern class $c_1(P)$ in the lattice $\BL^\vee$ and is
insensitive to torsion.  The sign in
\eqref{IUPAAA} is just a convention that will make the duality
formulas in Section \ref{Modular} more elegant. 

Dually, we also introduce a harmonic two-form $\upbeta$,
\begin{equation}
\upbeta \,\in\, \CH^2(M) \,\simeq\, \BR^{b_1}\,,
\end{equation}
with coupling 
\begin{equation}\label{UPBETAA}
{\bf I}_\upbeta(A) \,=\, \frac{1}{2\pi e^2} \int_M \*\upbeta\^F_A\,.
\end{equation}
The harmonic condition ensures that $\*\upbeta$ is a closed one-form,
so that the value of ${\bf I}_\upbeta(A)$ also depends only on the Chern class of
the bundle $P$.

We will eventually identify $\upalpha$ and $\upbeta$ with the same
parameters which we introduced for the periodic scalar field in
Section \ref{Scalar}.  However, the defining conditions on $\upalpha$
and $\upbeta$ are now reversed.  In Section
\ref{Scalar}, the one-form $\upalpha$ was required to be harmonic and
$\upbeta$ was an arbitrary closed two-form, whereas
here $\upbeta$ is harmonic and $\upalpha$ is an arbitrary closed
one-form.

Including the parameters $\upalpha$ and $\upbeta$, the Maxwell action
on $M$ becomes 
\begin{equation}\label{ITOTA}
\begin{aligned}
{\bf I}_{\rm tot}(A) \,&=\, {\bf I}_0(A) \,+\, {\bf I}_\upalpha(A)
\,+\, {\bf I}_\upbeta(A)\,,\\
&=\, {\bf I}_0(A) \,+\, \frac{1}{2\pi i}\int_M
\big(-\!\upalpha \,+\, \frac{i}{e^2}\,\*\upbeta\big)\^F_A\,.
\end{aligned}
\end{equation}
Without loss, I select a harmonic representative for $\upalpha$ and
introduce the complex harmonic one-form appearing holomorphically in
\eqref{ITOTA},
\begin{equation}\label{AZETA}
\upzeta \,=\, -\upalpha \,+\, \frac{i}{e^2}\,\*\upbeta 
\,\in\,\CH^1_\BC(M)\,.
\end{equation}
Just as for the harmonic two-form $\upgamma$ in \eqref{BZETA},
the harmonic one-form $\upzeta$ is invariant under the combined scale
transformations in \eqref{SCALEg} and \eqref{SCALEe}.  As one can
easily check, the complex harmonic forms $\upgamma$ and $\upzeta$
are related by 
\begin{equation}\label{THPARMS}
\upgamma \,=\, -i \, e^2 \, \*\upzeta\,\in\,\CH^2_\BC(M)\,.
\end{equation}

The total Maxwell action in \eqref{ITOTA} can then be written
concisely in terms of the $L^2$ and intersection pairings, 
\begin{equation}\label{TOTMC}
{\bf I}_{\rm tot}(A) \,=\, \frac{1}{4\pi e^2}\left(F_A, F_A\right)
\,+\, \frac{1}{2\pi i} \left\langle\upzeta, F_A\right\rangle.
\end{equation}

\noindent{\sl Abelian Duality at Level $k$}\medskip

As a special feature of abelian gauge theory in three dimensions, we
can add to the Maxwell action \eqref{TOTMC} a 
Chern-Simons term proportional to 
\begin{equation}\label{CSONE}
\SC\SS(A) \,=\, \frac{1}{2\pi}\int_M A\^dA\,.
\end{equation}
We follow the standard practice in writing the Chern-Simons functional
with respect to a local trivialization for the bundle $P$.  However, 
because $A$ can be a connection on a non-trivial $U(1)$-bundle $P$
over $M$, the global meaning of the trivialized form
\eqref{CSONE} of the Chern-Simons functional may be unclear.  

For an alternative presentation, one can always choose a four-manifold
$X$ such that $X$ bounds $M$ and the $U(1)$-bundle $P$ extends from
$M$ to $X$.  The existence of $X$ relies upon the vanishing of
$H_3(BU(1))$ and is discussed thoroughly in \cite{Dijkgraaf:90}.
Given $X$, the Chern-Simons functional on $M$ can be rewritten in the
gauge-invariant fashion 
\begin{equation}\label{CSTWO}
\SC\SS(A) \,=\, \frac{1}{2\pi} \int_X F_A \^
F_A\,,\qquad\qquad \partial X \,=\, M\,.
\end{equation}
Since ${F_A/2\pi}$ is an integral two-form on any closed manifold, and
since the intersection pairing is also integral, the global expression for the
Chern-Simons functional in \eqref{CSTWO} shows that the value of 
$\SC\SS(A)$ is well-defined modulo $2\pi$.

If we wish, we can then extend the Maxwell action \eqref{TOTMC} on $M$
to a Maxwell-Chern-Simons action at level ${k \in \BZ}$,
\begin{equation}
{\bf I}_{\rm MCS}(A) \,=\,  \frac{1}{4\pi e^2}\left(F_A, F_A\right)
\,+\, \frac{1}{2\pi i} \left\langle\upzeta, F_A\right\rangle \,-\, i
\, k \, \SC\SS(A)\,.
\end{equation}

Just as the parameter $\upzeta$ is naturally related under duality to
the parameter $\upgamma$ for the periodic scalar field $\phi$, one can ask
about the dual interpretation for the Chern-Simons level $k$.
One standard answer to this question would be to say that the Chern-Simons
level has no simple, local description in terms of the periodic scalar
field.  Strictly speaking, this answer is correct, but it is
unsatisfying.  Another standard answer, at least when ${M = \BR^3}$,
would be to say that Maxwell-Chern-Simons theory at level $k$ is equivalent
\cite{Deser:1984kw,Karlhede:1986qd} to the  `self-dual' model
\cite{Townsend:1984} of a massive, non-gauge-invariant Proca vector
field with Chern-Simons term.  Strictly
speaking, this answer is also correct, but it is not correct for a
general three-manifold.

I will discuss elsewhere a better, more global
answer to the question  ``What is the dual of the
Chern-Simons level?''  The answer turns out to be most 
clear with the Hamiltonian formalism developed in \cite{BeasleyII}.  For the
present, I just set ${k=0}$ and work with only the pure Maxwell theory
on $M$. 

\subsection{Computing the Partition Function}

The Maxwell partition function on $M$ can be evaluated in a manner
very similar to the evaluation of the scalar partition in
Section \ref{ScalarP}.  So I will be relatively brief.

The Maxwell partition function admits the formal path integral
presentation 
\begin{equation}\label{ZMDUAL}
Z^\vee_M(\upzeta) \,=\, \sum_{c_1(P) \in H^2(M;\BZ)}
\frac{1}{\Vol(\CG(P))} \int_{\CA(P)}\!\!\CD A \, \exp{\!\left[-{\bf
      I}_{\rm tot}(A)\right]}\,.
\end{equation}
Evidently, the partition function involves both a sum over the topological
type of the principal $U(1)$-bundle $P$ as well as an integral over the affine
space $\CA(P)$ of all connections on $P$.  Due to the gauge invariance
of the Maxwell action, we divide the path integral by the volume
of the group $\CG(P)$ of gauge transformations on $P$.  Geometrically,
$\CG(P)$ can be identified with the group of maps from $M$ to $U(1)$, 
\begin{equation}
\CG(P) \,=\, \Map\big(M, U(1)\big)\,,
\end{equation}
acting on $P$ by bundle automorphisms.

Again, the most delicate aspect of our computation will be to fix the
normalization of the partition function, for which we must be
precise about the meaning of the measure $\CD A$ on $\CA(P)$.  

\medskip\noindent{\sl More about the Maxwell Measure}\medskip

Let $P_0$ be the trivial $U(1)$-bundle over $M$.  For every other
bundle $P$, the space $\CA(P)$ can be identified with $\CA(P_0)$ as
soon as we pick a basepoint in $\CA(P)$, which will correspond
geometrically to a fiducial connection on $P$.  In close analogy to
the choice of the fiducial harmonic map in Section \ref{Scalar}, we
take the fiducial connection ${\widehat{A}_P\in\CA(P)}$ to possess
harmonic curvature 
\begin{equation}\label{HARMFAII}
F_{\widehat{A}_P} \,=\, 2 \pi \lambda\,,\qquad\qquad \lambda \,\in\,
\BL^\vee = H^2(M;\BZ)_{\rm free}\,,
\end{equation}
as well as trivial holonomy around a fixed set of curves ${C \subset
  M}$ which represent generators for $H_1(M;\BR)$.

The arbitrary connection ${A \in \CA(P)}$ can then be expressed as a sum 
\begin{equation}\label{SUMA}
A \,=\, \widehat{A}_P \,+\, \eta\,,\qquad\qquad \eta\,\in\,\CA(P_0)\,,
\end{equation}
where $\eta$ is a connection on the trivial bundle.  The correspondence
between $A$ and $\eta$ in \eqref{SUMA} provides the requisite
identification ${\CA(P) \simeq \CA(P_0)}$ for each $U(1)$-bundle $P$.
Given this identification, we need only describe the measure $\CD A$
for connections on the trivial bundle over $M$.

The fiducial connection on the trivial bundle $P_0$ is flat, from
which we obtain a trivialization of $P_0$.  We may thus regard
connections on $P_0$ as ordinary one-forms on $M$.  Following the
same philosophy from Section \ref{Scalar}, we characterize
the measure $\CD A$ on ${\CA(P_0) \simeq \Omega^1(M)}$ as the
Riemannian measure induced from the $L^2$-norm 
\begin{equation}\label{BIGA}
||\delta A||^2_{\CA(P_0)} \,=\, \frac{e^2}{\left(2\pi\right)^2}\int_M
\delta A \^\* \delta A\,,\qquad\qquad \delta A \,\in\,\Omega^1(M)\,.
\end{equation}
Like the corresponding expression in \eqref{BIGM}, the factor of
$e^2$ in \eqref{BIGA} is dictated by invariance under the scaling in
\eqref{SCALEg} and \eqref{SCALEe}, and the
factors of $2\pi$ will prove to be a later numerical convenience.  By
construction, the measure $\CD A$ is invariant under translations in
$\CA(P_0)$.  Hence $\CD A$ does not depend upon the choice of fiducial
connection used to identify ${\CA(P_0) \simeq \Omega^1(M)}$.

Lastly,  to describe the volume factor appearing in
\eqref{ZMDUAL}, we must introduce a measure on the group $\CG(P)$ of gauge
transformations.  As the group of maps to $U(1)$, the Lie 
algebra of $\CG(P)$ is simply the linear space $\Omega^0(M)$ of functions on
$M$, with trivial Lie bracket,
\begin{equation}\label{LIEG}
\Lie\!\big(\CG(P)\big) \,=\, \Omega^0(M)\,.
\end{equation}
We have already introduced a suitable Riemannian metric on
$\Omega^0(M)$ in \eqref{BIGM}.  We extend this metric in a
translation-invariant fashion over $\CG(P)$, and we take
$\Vol(\CG(P))$ to be the formal Riemannian volume.  This
volume is independent of the bundle $P$.

\medskip\noindent{\sl A Sum over Fluxes}\medskip

Once we substitute for $A$ as in \eqref{SUMA}, the Maxwell partition
function can be rewritten as a path integral over connections on the
trivial bundle $P_0$ alone,
\begin{equation}\label{SUMZV}
Z^\vee_M\big(\upzeta\big) \,=\, \sum_{c_1(P) \in H^2(M;\BZ)}
\frac{1}{\Vol(\CG(P_0))} \int_{\CA(P_0)}\!\!\CD\eta \, \exp{\!\left[-{\bf
      I}_{\rm tot}\!\left(\widehat{A}_P + \eta\right)\right]}\,.
\end{equation}
In terms of the fiducial connection $\widehat{A}_P$ and the one-form
$\eta$, the Maxwell action becomes
\begin{equation}\label{IAEN}
{\bf I}_{\rm tot}\!\left(\widehat{A}_P + \eta\right) \,=\, \frac{1}{4\pi
  e^2}\left(F_{\widehat{A}_P} + d\eta, F_{\widehat{A}_P} + d\eta\right)
\,+\, \frac{1}{2\pi i}\left(\upzeta, F_{\widehat{A}_P} +
  d\eta\right),
\end{equation}
or more explicitly,
\begin{equation}\label{IAENII}
{\bf I}_{\rm tot}\!\left(\widehat{A}_P + \eta\right) \,=\, \frac{1}{4\pi
  e^2}\int_M\left(2\pi\lambda\,+\, d\eta)\^\*(2\pi\lambda \,+\,
  d\eta\right) +\, \frac{1}{2\pi i}\int_M
\upzeta\^\!\left(2\pi\lambda \,+\, d\eta\right).
\end{equation}
In passing to \eqref{IAENII}, we recall the formula
for the harmonic curvature $F_{\widehat{A}_P}$ in \eqref{HARMFAII}.
Since $\lambda$ and $\upzeta$ are harmonic, all cross-terms which involve
$\lambda$ or $\upzeta$ together with $d\eta$ vanish, and
\begin{equation}\label{MWSA}
\begin{aligned}
{\bf I}_{\rm tot}\!\left(\widehat{A}_P + \eta\right) \,&=\,
\frac{1}{4\pi e^2} \int_M d\eta\^\*d\eta \,+\, \frac{\pi}{e^2}\int_M
\lambda\^\*\lambda \,-\, i\int_M\upzeta\^\lambda\,,\\
&=\, \frac{1}{4\pi e^2} \left(d\eta, d\eta\right) \,+\,
\frac{\pi}{e^2}\left(\lambda, \lambda\right) \,-\,i
\left\langle\upzeta, \lambda\right\rangle.
\end{aligned}
\end{equation}

With this description for the Maxwell action, the partition function in
\eqref{SUMZV} takes the more explicit form 
\begin{equation}\label{MWSAII}
\begin{aligned}
&Z^\vee_M\big(\upzeta\big)\,=\\
&\qquad {\Tor}_M\cdot\sum_{\lambda \in \BL^\vee}
\frac{1}{\Vol(\CG(P_0))} \int_{\CA(P_0)}\!\!\!\!\CD\eta \,
\exp{\!\left[-\frac{1}{4\pi e^2}\left(d\eta,d\eta\right) -
    \frac{\pi}{e^2}\left(\lambda,\lambda\right) +
    i\,\langle\upzeta,\lambda\rangle\right]}\,.
\end{aligned}
\end{equation}
Here $\Tor_M$ is the number of elements in the torsion subgroup of $H^2(M;\BZ)$,
\begin{equation}
{\Tor}_M \,=\, \left|H^2(M;\BZ)_{\rm tors}\right|.
\end{equation}
Since the Maxwell action is insensitive to torsion in
$c_1(P)$, the sum over $H^2(M;\BZ)$ in \eqref{SUMZV} reduces to a sum
over the quotient lattice $\BL^\vee$ in \eqref{MWSAII}.

Like the partition function \eqref{FACTZM} of the periodic
scalar field, the Maxwell partition function also factorizes,
\begin{equation}
Z^\vee_M(\upzeta) \,=\, \Delta^\vee_M \cdot \Theta^\vee_M(\upzeta)\,.
\end{equation}
Here $\Delta^\vee_M$ is given by a path integral over the affine
space $\CA(P_0)$,
\begin{equation}
\Delta^\vee_M \,=\, \Tor_M\cdot\frac{1}{\Vol(\CG(P_0))}
\int_{\CA(P_0)}\!\!\CD\eta \, \exp{\!\left[-\frac{1}{4\pi
      e^2}\left(d\eta,d\eta\right)\right]}\,,
\end{equation}
and $\Theta^\vee_M(\upzeta)$ is given by a sum over fluxes in the
quotient lattice ${\BL^\vee = H^2(M;\BZ)_{\rm free}}$,
\begin{equation}
\Theta^\vee_M(\upzeta) \,=\, \sum_{\lambda\in\BL^\vee}
\exp{\!\left[-\frac{\pi}{e^2}\left(\lambda,\lambda\right) +
    i\,\langle\upzeta,\lambda\rangle\right]}\,.
\end{equation}

The more interesting factor in the Maxwell partition function is 
$\Theta^\vee_M(\upzeta)$, which is yet another theta-function
attached to the three-manifold $M$.  Including our previous 
work from Section \ref{Scalar}, we now have a dual pair of lattices $\BL$
and $\BL^\vee$, as well as a pair of theta-functions
$\Theta_M^{}$ and $\Theta^\vee_M$.  As one might guess, and as we will
demonstrate explicitly in Section \ref{Modular}, $\Theta_M^{}$ and
$\Theta^\vee_M$ are related by a modular transformation.  Before we
discuss modular issues though, let us finish the computation of the
Maxwell partition function on $M$.

\medskip\noindent{\sl The Normalization of
  $Z^\vee_M(\upzeta)$}\medskip

To fix the absolute normalization of the Maxwell partition function,
which will depend upon the electric coupling $e^2$ and the Riemannian
metric $g$, we are left to evaluate the Gaussian path integral 
\begin{equation}\label{DELCM}
\Delta^\vee_M \,=\, \Tor_M\cdot\frac{1}{\Vol(\CG)}
\int_{\CA}\! \CD\eta \, \exp{\!\left[-\frac{1}{4\pi
      e^2}\left(d\eta,d\eta\right)\right]}\,.
\end{equation}
Here I abbreviate ${\CA \equiv \CA(P_0)}$ and ${\CG \equiv \CG(P_0)}$,
since we will only consider gauge theory on the trivial $U(1)$-bundle
$P_0$ for the remainder of the discussion. Also, to orient the reader,
I recall that the one-form $\eta$ in \eqref{DELCM} is effectively
identified with the gauge field $A$ after the bundle $P_0$ has been
trivialized.

The computation of $\Delta^\vee_M$ is slightly more delicate than
the analogous computation for the periodic scalar field, due to the gauge 
symmetry in the current problem.  Because of the gauge symmetry, the
argument of the exponential in \eqref{DELCM} vanishes for any
${\eta\in\Omega^1(M)}$ of the form ${\eta = d\varphi}$, with
${\varphi\in\Omega^0(M)}$.  Intrinsically, $\varphi$ can be
interpreted as element in the Lie algebra of the group
$\CG$, and we are simply observing that the Maxwell action is degenerate along
orbits of $\CG$.

To account for the degeneracy of the integrand in \eqref{DELCM}, we
employ the standard BRST technique to fix the gauge symmetry. 
We cannot possibly fix the action for the full group 
$\CG$ of all gauge transformations, since any gauge
transformation generated by a constant function ${\varphi_0\in\BR}$
acts everywhere trivially on $\CA$.  Instead, we pick a point ${p \in
  M}$, and we consider the subgroup ${\CG_p \subset \CG}$ of gauge
transformations which are {\em based} at $p$.  An alternative
treatment would involve the introduction of BRST 
ghosts-for-ghosts to deal with the constant gauge transformations, but
I believe that working with the based gauge group is conceptually
simpler for this example.

By definition, elements in $\CG_p$ are gauge transformations which are
the identity at the point $p$, and elements in the Lie algebra of $\CG_p$
are functions ${\varphi\in\Omega^0(M)}$ which vanish at $p$,
\begin{equation}\label{BASEDP}
\varphi(p) \,=\, 0\,,\qquad\qquad \varphi \,\in\, \Lie(\CG_p)\,.
\end{equation}
Due to the condition in \eqref{BASEDP}, the identity is the only
constant gauge transformation in $\CG_p$, and the quotient of $\CG$ by
$\CG_p$ is the group 
\begin{equation}
\CG/\CG_p\,=\, U(1)\,,
\end{equation}
acting globally by constant gauge transformations on $M$.  Rather than
attempt to fix a gauge for $\CG$, we instead fix a gauge for the 
slightly smaller, based group $\CG_p$.

As usual in the BRST procedure, we introduce additional fields $c$,
$\bar c$, and $h$, all valued in the Lie algebra of $\CG_p$.  Thus
$(c,\bar c, h)$ are functions on $M$ which vanish at $p$, 
\begin{equation}\label{CONSTR}
c(p) = \bar c(p) = h(p) = 0\,,\qquad\qquad c, \bar c, h\,\in\,\Omega^0(M)\,.
\end{equation} 
By assumption, $c$ and $\bar c$ are anti-commuting, Grassmann scalar fields,
and $h$ is a commuting scalar field.  If one wishes, the vanishing
constraint in \eqref{CONSTR} amounts to the insertion of a local
operator $\CO(p)$, whose role is to absorb the zero-modes of $(c,\bar c,
h)$ which would otherwise be present in the BRST path integral.

To achieve the most elegant geometric formulation of the BRST
procedure, I will depart somewhat from custom and introduce an extra
bosonic field $u$, which will be an element in the based group $\CG_p$.
Equivalently, $u$ is a sigma model map from $M$ to $U(1)$, satisfying 
\begin{equation}
u:M\,\rightarrow\, U(1)\,,\qquad\qquad u(p) \,=\, 1\,.
\end{equation}
Together, the pair $(u,h)$ describes the cotangent bundle of the group
$\CG_p$,
\begin{equation}
T^*\CG_p \,\simeq\, \CG_p\times\Lie\!\left(\CG_p\right)\,,
\end{equation}
and the anti-commuting
scalars $(c,\bar c)$ can be interpreted as one-forms on
$T^*\CG_p$.\footnote{The bar on $\bar c$ does not indicate complex
  conjugation.  The notation is traditional.}

The nilpotent BRST operator $Q$ acts infinitesimally on the set of fields
$\left(\eta,h,u,c,\bar c\right)$ according to 
\begin{equation}\label{BIGQ}
\begin{matrix}
\begin{aligned}
\delta \eta \,&=\, \frac{i}{2\pi}\,dc\,,\\
\delta h \,&=\, 0\,,\\ 
\delta u\,&=\, 0\,.
\end{aligned} \quad & \quad
\begin{aligned}
\delta c\,&=\, 0\,,\\
\delta\bar c \,&=\, h\,,\\ \\
\end{aligned}
\end{matrix}
\end{equation}
Manifestly ${Q^2 = 0}$, and $Q$ annihilates the Maxwell action in
\eqref{DELCM} by virtue of gauge invariance.  

Using the BRST charge $Q$, we produce a gauge-fixing action ${\bf
  I}_{\rm g.f.}$ appropriate for harmonic gauge ${d^\dagger\eta = 0}$,
\begin{equation}
{\bf I}_{\rm g.f.} \,=\, \int_M \left\{Q, {\bf
    V}\right\}\,,\qquad\qquad {\bf V} \,=\, \bar c
\^\*\!\left(\frac{e^6}{4\pi} \, h 
  \,+\, i \, \frac{e^2}{2\pi} \, d^\dagger\eta^u\right)\,.
\end{equation}
Here ${\eta^u = \eta + i\, u^{-1} du}$ is the image of the one-form
$\eta$ under a gauge transformation by $u$.  The
various factors of $e^2$ ensure invariance 
under the scaling in \eqref{SCALEg} and \eqref{SCALEe}, and the
factors of $2\pi$ are a numerical convenience, related to all the
other factors of $2\pi$ that are floating around!  Explicitly from \eqref{BIGQ},
\begin{equation}\label{IGFII}
{\bf I}_{\rm g.f.} \,=\, \int_M \left(\frac{e^6}{4\pi} \, h\^\*h
  \,+\, i \, \frac{e^2}{2\pi} \, h 
  \^\*d^\dagger\eta^u \,+\, \frac{e^2}{\left(2\pi\right)^2}\, \bar
  c\^\*\triangle_0 c\right),
\end{equation}
where ${\triangle_0 = d^\dagger d}$ is the scalar Laplacian on $M$.

The essence of the BRST procedure amounts to an amusing way to
rewrite unity,
\begin{equation}\label{FPFG}
1 \,=\, \int_{T^*\CG_p} \CD u \, \CD h \, \CD c \, \CD \bar c \, 
\exp{\!\big[-{\bf I}_{\rm g.f.}\big]}\,.
\end{equation}
A special feature of the path integral in \eqref{FPFG} is the pairing of the
bosonic measure ${\CD u \, \CD h}$ with the fermionic measure ${\CD c
  \, \CD \bar c}$.  Each of $\CD u$, $\CD h$, $\CD c$, and $\CD \bar
c$ can be defined once a metric on the Lie algebra of $\CG_p$ is
chosen.  Provided that we make the same choice throughout, this choice
does not matter, due to the familiar cancellation of Jacobians between
bosons and fermions.  But to make a definite choice, we use the
scale-invariant version of the $L^2$-norm in \eqref{BIGM}.

Otherwise, the core of the BRST identity \eqref{FPFG} is not so
much the appearance of the constant `$1$' on the left-hand side of the
identity as the independence of the right-hand side on the 
one-form $\eta$ which enters the gauge-fixing action ${\bf I}_{\rm
  g.f.}$ in \eqref{IGFII}.  The latter property is really a property
of harmonic gauge:~for any one-form $\eta$, a gauge transformation by
a unique ${u \in \CG_p}$ exists so that ${d^\dagger\eta^u = 0}$.
Given this statement, which follows from standard Hodge theory, the
path integral over $\CG_p$ washes out all dependence
on $\eta$ in the integrand of \eqref{FPFG}.

Using the BRST identity, we enlarge the path integral which describes
$\Delta^\vee_M$ in \eqref{DELCM} to a path integral over the product
${\CA \times T^*\CG_p}$, 
\begin{equation}
\begin{aligned}
\Delta^\vee_M \,=\, \Tor_M\cdot\frac{1}{\Vol(\CG)}
&\int_{\CA \times T^*\CG_p}\! \CD\eta \, \CD u \, \CD h \, \CD c \, \CD
\bar c \,\, \exp{\!\left[-\frac{1}{4\pi
      e^2}\left(d\eta,d\eta\right)\right]}\,\times\,\\
&\times\,\exp{\!\left[-\frac{e^6}{4\pi}\left(h, h\right) -
    i\,\frac{e^2}{2\pi}\left(h, d^\dagger\eta^u\right) -
    \frac{e^2}{\left(2\pi\right)^2}\left(\bar c,\triangle_0 c\right)\right]}\,.
\end{aligned}
\end{equation}
The Gaussian integral over the auxiliary scalar $h$ can be evaluated
immediately, after which the normalization factor becomes 
\begin{equation}\label{DELCMII}
\begin{aligned}
\Delta^\vee_M \,=\, \Tor_M\cdot\frac{1}{\Vol(\CG)}
&\int_{\CA \times \CG_p}\! \CD\eta \, \CD u \, \CD c \, \CD
\bar c \,\, \exp{\!\left[-\frac{1}{4\pi
      e^2}\left(d\eta,d\eta\right)\right]}\,\times\\ 
&\times\,\exp{\!\left[-\frac{1}{4\pi e^2}\left(d^\dagger\eta^u,
      d^\dagger\eta^u\right) - \frac{e^2}{\left(2\pi\right)^2}\left(\bar
      c,\triangle_0 c\right)\right]}\,. 
\end{aligned}
\end{equation}

To deal with the appearance of $u$ in the integrand of
\eqref{DELCMII}, we note trivially 
\begin{equation}
\left(d\eta,d\eta\right) \,=\, \left(d\eta^u,d\eta^u\right),
\end{equation}
due to gauge-invariance of the Maxwell action.  Gauge-invariance for
the measure on $\CA$ similarly implies ${\CD\eta = \CD\eta^u}$.  As a
result, $\eta$ can be replaced by its gauge transform
$\eta^u$ in \eqref{DELCMII}, 
\begin{equation}\label{DELCMIII}
\begin{aligned}
\Delta^\vee_M \,=\, \Tor_M\cdot\frac{1}{\Vol(\CG)}
&\int_{\CA \times \CG_p}\! \CD u \, \CD\eta^u \, \CD c \, \CD
\bar c \,\, \exp{\!\left[-\frac{1}{4\pi
      e^2}\left(d\eta^u, d\eta^u\right)\right]}\,\times\\ 
&\times\,\exp{\!\left[-\frac{1}{4\pi e^2}\left(d^\dagger\eta^u,
      d^\dagger\eta^u\right) - \frac{e^2}{\left(2\pi\right)^2}\left(\bar
      c,\triangle_0 c\right)\right]}\,.
\end{aligned}
\end{equation}
After a change-of-variables from $\eta^u$ back to $\eta$, the
auxiliary field ${u \in \CG_p}$ decouples from the integrand in
\eqref{DELCMIII}, so that 
\begin{equation}\label{DELCMIV}
\begin{aligned}
\Delta^\vee_M \,=\, \Tor_M\cdot\frac{1}{\Vol(\CG)}
&\int_{\CA \times \CG_p}\! \CD u \, \CD\eta \, \CD c \, \CD
\bar c \,\, \exp{\!\left[-\frac{1}{4\pi
      e^2}\left(d\eta, d\eta\right)\right]}\,\times\\ 
&\times\,\exp{\!\left[-\frac{1}{4\pi e^2}\left(d^\dagger\eta,
      d^\dagger\eta\right) - \frac{e^2}{\left(2\pi\right)^2}\left(\bar
      c,\triangle_0 c\right)\right]}\,.
\end{aligned}
\end{equation}
Since $u$ appears nowhere in the integrand of \eqref{DELCMIV}, the
path integral over $u$ just contributes a factor of the group volume
$\Vol(\CG_p)$,
\begin{equation}\label{DELCMV}
\Delta^\vee_M \,=\, \Tor_M\cdot\frac{\Vol(\CG_p)}{\Vol(\CG)}\int_\CA\!
\CD\eta \, \CD c \, \CD \bar c \,\,\exp{\!\left[-\frac{1}{4\pi
      e^2}\left(\eta,\triangle_1\eta\right) -
    \frac{e^2}{\left(2\pi\right)^2}\left(\bar c,\triangle_0 c\right)\right]}\,.
\end{equation}
In passing from \eqref{DELCMIV} to \eqref{DELCMV}, we also integrate
by parts to produce the de Rham Laplacian ${\triangle_1 = d^\dagger d
  + d d^\dagger}$ acting on the one-form $\eta$.

Although both $\CG_p$ and $\CG$ have infinite dimension, the quotient
${\CG/\CG_p = U(1)}$ has finite dimension, and the ratio of volumes in
\eqref{DELCMV} is well-defined,
\begin{equation}\label{VOLSTB}
\frac{\Vol(\CG_p)}{\Vol(\CG)} \,=\, \frac{1}{\Vol\!\left(U(1)\right)}\,.
\end{equation}
Because $U(1)$ acts by constant gauge transformations, $U(1)$ is the
stabilizer at all points in $\CA$.  As usual, the factor in
\eqref{VOLSTB} implies that the partition function on $M$ is divided
by the volume of the stabilizer.  See for instance \S $2.2$ in
\cite{Witten:1991we} for a related discussion of the role of
stabilizers in $\CG$ and the normalization of the gauge theory
partition function.

The Gaussian path integral over $\eta$, $c$, and $\bar c$ can be
formally evaluated by expanding each field in an orthonormal basis of
eigenmodes for the Laplacian, exactly as we did previously for the
periodic scalar field in \eqref{ONDME}.  With care for factors of
$e^2$, one finds 
\begin{equation}\label{INCDAE}
\int_\CA\! \CD\eta \, \CD c \, \CD \bar c \, \exp{\!\left[-\frac{1}{4\pi
      e^2}\left(\eta,\triangle_1\eta\right) -
    \frac{e^2}{\left(2\pi\right)^2}\left(\bar c,\triangle_0
      c\right)\right]} =
\frac{\det'\!\left(\triangle_0/e^4\right)}{\sqrt{\det'\!\left(\triangle_1/e^4\right)}}\cdot\Vol\!\left(\SJ_M\right). 
\end{equation}
The functional determinants of the respective scalar and vector
Laplacians $\triangle_{0,1}$ arise from the Gaussian integrals over
non-harmonic modes of $\left(\eta,c,\bar c\right)$, and the volume of
the torus $\SJ_M$ in \eqref{JACM} arises from the integral over the
remaining harmonic modes of $\eta$.  Exactly as in \eqref{DMFIN}, the
factor of $1/e^4$ in each functional determinant is required by
invariance under the scaling in \eqref{SCALEg} and \eqref{SCALEe} and
is a consequence of the coupling-dependence in the metrics on
$\Omega^0(M)$ in \eqref{BIGM} and $\Omega^1(M)$ in \eqref{BIGA}.

In total, the results in \eqref{DELCMV}, \eqref{VOLSTB}, and
\eqref{INCDAE} imply 
\begin{equation}\label{INCDAIEII}
\Delta^\vee_M \,=\, \Tor_M \cdot
\frac{\Vol\!\left(\SJ_M\right)}{\Vol\!\left(U(1)\right)} \cdot
\frac{\det'\!\left(\triangle_0/e^4\right)}{\sqrt{\det'\!\left(\triangle_1/e^4\right)}}\,.
\end{equation}

Both the volume of ${U(1)\subset \CG}$ and the volume of $\SJ_M$ are to be
evaluated using the metrics induced from the coupling-dependent $L^2$-norms in
\eqref{BIGM} and \eqref{BIGA}.  With respect to \eqref{BIGM}, the norm-square
of the constant function `$1$' is 
\begin{equation}
||1||^2_{\Omega^0(M)} \,=\,
\frac{\left(e^2\ell\right)^3}{\left(2\pi\right)^2}\,,
\end{equation}
from which we obtain 
\begin{equation}\label{VOLUONE}
\begin{aligned}
\Vol\!\left(U(1)\right) \,=\, 2\pi\,||1||_{\Omega^0(M)} \,=\,
(e^2\ell)^{3/2}\,.
\end{aligned}
\end{equation}
The same factor appears in \eqref{DMFIN}, for exactly the same reason.

To determine the volume of $\SJ_M$, we recall that
$\SJ_M$ is concretely the quotient 
\begin{equation}
\SJ_M \,=\, H^1(M;\BR)/2\pi\BL\,,\qquad\qquad \BL \,=\,
H^1(M;\BZ)\,.
\end{equation}
Let $\left\{\Fe_1,\cdots,\Fe_{b_1}\right\}$ be a basis of integral
generators for $\BL$,
\begin{equation}\label{INTB}
\BL \,\simeq\, \BZ \Fe_1 \oplus \cdots \oplus \BZ \Fe_{b_1}\,,
\end{equation}
so that $\SJ_M$ becomes isomorphic to 
\begin{equation}\label{CONJM}
\SJ_M \,\simeq\, \BR^{b_1}/2\pi\BZ^{b_1}\,.
\end{equation}
Associated to the integral basis in \eqref{INTB} is the matrix of
$L^2$ inner-products 
\begin{equation}\label{BIGSQ}
\SQ_{j k} \,=\, \left(\Fe_j, \Fe_k\right) \,=\, \int_M
\Fe_j\^\*\Fe_k\,,\qquad\qquad j,k = 1,\ldots,b_1\,,
\end{equation}
where we implicitly use the embedding ${\BL \subset \CH^1(M)}$ to
identify the generators of $\BL$ with harmonic one-forms
on $M$.  Manifestly, $\SQ$ is a symmetric, positive-definite matrix, and basic
linear algebra implies that the volume of $\SJ_M$ in \eqref{CONJM} is
proportional to the square-root of the determinant of $\SQ$,
\begin{equation}\label{VOLJM}
\begin{aligned}
\Vol\!\left(\SJ_M\right) \,&=\, \left(2\pi\right)^{b_1} \cdot
\left(\frac{e}{2\pi}\right)^{b_1}\sqrt{\det\SQ}\,,\\
&=\, e^{b_1} \sqrt{\det\SQ}\,.
\end{aligned}
\end{equation}
The extra factor of $\left(e/2\pi\right)^{b_1}$ in the first line of
\eqref{VOLJM} occurs due to the corresponding factor in the
scale-invariant norm on $\Omega^1(M)$ in \eqref{BIGA}.

Finally, we extract factors of $e^2$ from the functional determinants
in \eqref{INCDAIEII} using the zeta-function relation in
\eqref{SCALED}.  According to the general formula \eqref{ZETAZ} for the value of
the zeta-function at ${s=0}$,
\begin{equation}
\zeta_{\triangle_0}(0) \,=\, -1\,,\qquad\qquad \zeta_{\triangle_1}(0)
\,=\, -b_1\,,
\end{equation}
so again,
\begin{equation}\label{SCELDD}
{\det}'\!\left(\triangle_0/e^4\right) \,=\, e^4\cdot
{\det}'\!\left(\triangle_0\right),\qquad\quad
{\det}'\!\left(\triangle_1/e^4\right) \,=\, e^{4 b_1}\cdot
{\det}'\!\left(\triangle_1\right).
\end{equation}

We use the formulas in \eqref{VOLUONE}, \eqref{VOLJM}, and
\eqref{SCELDD} to simplify our result in
\eqref{INCDAIEII},
\begin{equation}\label{FDELVM}
\Delta^\vee_M \,=\, \Tor_M \cdot \frac{e^{1-b_1}}{\ell^{3/2}}
\sqrt{\det\SQ} \cdot
\frac{\det'\!\left(\triangle_0\right)}{\sqrt{\det'\!\left(\triangle_1\right)}}\,.
\end{equation}
The overall dependence of $\Delta^\vee_M$ on the electric coupling as
$e^{1-b_1}$ can also be understood more physically (and perhaps more
simply) by counting modes of the gauge field $A$ modulo gauge
equivalence.  The latter perspective is taken for the computations in
\cite{BrodaWG} and \cite{WittenGF}.

\medskip\noindent{\sl The Final Result}\medskip

In summary, we have determined the Maxwell partition function on $M$
to be 
\begin{equation}\label{FINMAXZ}
\begin{aligned}
Z^\vee_M(\upzeta) \,&=\,  \Delta^\vee_M \cdot
\Theta^\vee_M(\upzeta)\,,\qquad\qquad \upzeta \,\in\,
\CH^1_\BC(M)\,,\\
&=\, \Tor_M \cdot \frac{e^{1-b_1}}{\ell^{3/2}}
\sqrt{\det\SQ} \cdot
\frac{\det'\!\left(\triangle_0\right)}{\sqrt{\det'\!\left(\triangle_1\right)}}
\cdot \sum_{\lambda\in\BL^\vee}
\exp{\!\left[-\frac{\pi}{e^2}\left(\lambda,\lambda\right) +
    i\,\langle\upzeta,\lambda\rangle\right]}\,.
\end{aligned}
\end{equation}
As a reminder, $\Tor_M$ is the order of the torsion subgroup in $H^2(M;\BZ)$,
\begin{equation}
\Tor_M \,=\, \big|H^2(M;\BZ)_{\rm tors}\big|\,,
\end{equation}
and $\SQ$ is the matrix of inner-products for an integral basis
$\left\{\Fe_1,\cdots,\Fe_{b_1}\right\}$ of $H^1(M;\BZ)$,
\begin{equation}
\SQ_{j k} \,=\, \left(\Fe_j, \Fe_k\right) \,=\, \int_M
\Fe_j\^\*\Fe_k\,,\qquad\qquad j,k = 1,\ldots,b_1\,.
\end{equation}

\section{Modularity, Duality, and All That}\label{Modular}

Having evaluated the respective scalar and Maxwell partition functions,
we now compare these results.  Both $Z_M(\upgamma)$ and
$Z^\vee_M(\upzeta)$ factorize,
\begin{equation}\label{BOTHZS}
\begin{aligned}
Z_M(\upgamma) \,&=\, \Delta_M \cdot
\Theta_M(\upgamma)\,,\qquad\qquad \upgamma\,\in\,\CH^2_\BC(M)\,,\\
Z^\vee_M(\upzeta) \,&=\,  \Delta^\vee_M \cdot
\Theta^\vee_M(\upzeta)\,,\qquad\qquad \upzeta\,\in\,\CH^1_\BC(M)\,,
\end{aligned}
\end{equation}
and we will start by comparing the respective theta-functions
$\Theta_M^{}(\upgamma)$ and $\Theta^\vee_M(\upzeta)$ associated to the
three-manifold $M$.  See \cite{Mumford:83} or Ch.\,$2$ in
\cite{Griffiths:78} for an introduction to the geometry of
theta-functions, the basics of which will be useful here.

\subsection{A Theta-Function for Three-Manifolds}

The hallmark of any theta-function is quasi-periodic behavior under
integral shifts in the argument, and both $\Theta_M^{}(\upgamma)$ and
$\Theta^\vee_M(\upzeta)$ are easily seen to be quasi-periodic with
respect to shifts in the variables $\upgamma$ and
$\upzeta$.  

For convenience, I focus on $\Theta_M(\upgamma)$, given by the lattice sum
\begin{equation}\label{BIGXII}
\Theta_M(\upgamma) \,=\, \sum_{\omega\in\BL}\,\exp{\!\left[-\pi
    e^2 \left(\omega,\,\omega\right) \,+\, i
    \left\langle\omega,\upgamma\right\rangle\right]}\,,\qquad\qquad
\BL \,=\, H^1(M;\BZ)\,.
\end{equation}
Recall that $\BL$ is dual to the quotient lattice ${\BL^\vee =
  H^2(M;\BZ)_{\rm free}}$.  If ${\nu \in \BL^\vee}$ is such an
integral two-form, then manifestly 
\begin{equation}\label{PERD}
\Theta_M(\upgamma \,+\, 2 \pi \nu) \,=\,
\Theta_M(\upgamma)\,,\qquad\qquad \nu \,\in\,\BL^\vee\,.
\end{equation}
Similarly, if ${\mu \in \BL}$ is an integral one-form, then  
\begin{equation}\label{SHFTL}
\Theta_M(\upgamma \,+\, 2\pi i \, e^2\, \*\mu) \,=\, \exp{\!\left[\pi 
    e^2
    \left(\mu,\mu\right) \,-\, i\,
    \langle\mu,\upgamma\rangle\right]}\cdot\Theta_M(\upgamma)\,,\qquad\quad
\mu \,\in\,\BL\,.
\end{equation}
The transformation formula in \eqref{SHFTL} follows by a standard
calculation from the lattice sum in \eqref{BIGXII},
\begin{equation}\label{QUASIP}
\begin{aligned}
&\Theta_M(\upgamma \,+\, 2\pi i \, e^2\, \*\mu) \,=\,
\sum_{\omega\in\BL}\,\exp{\!\left[-\pi 
    e^2 \left(\omega,\,\omega\right) \,+\, i
    \left\langle\omega,\upgamma\right\rangle - 2\pi e^2
    \left(\omega,\mu\right)\right]}\,,\\
&\qquad\,=\,\exp{\!\left[\pi e^2 \left(\mu,\mu\right)\right]}\cdot
\sum_{\omega\in\BL}\,\exp{\!\left[-\pi  
    e^2 \left(\omega + \mu,\,\omega + \mu\right) \,+\, i
    \left\langle\omega,\upgamma\right\rangle\right]}\,,\\
&\qquad\,=\, \exp{\!\left[\pi e^2 \left(\mu,\mu\right) \,-\, i \,
    \langle\mu,\upgamma\rangle\right]}\cdot
\sum_{\omega'\in\BL}\,\exp{\!\left[-\pi  
    e^2 \left(\omega',\,\omega'\right) \,+\, i
    \left\langle\omega',\upgamma\right\rangle\right]}\,,\\
&\qquad\,=\, \exp{\!\left[\pi e^2 \left(\mu,\mu\right) \,-\, i \,
    \langle\mu,\upgamma\rangle\right]} \cdot \Theta_M(\upgamma)\,.
\end{aligned}
\end{equation}
In passing from the second to the third line of \eqref{QUASIP}, I
shift the summand to $\omega' = \omega + \mu$, since ${\mu\in\BL}$
is also integral.  

Together, the transformation laws in \eqref{PERD} and \eqref{SHFTL}
show that $\Theta_M(\upgamma)$ is quasi-periodic with
respect to the lattice 
\begin{equation}
2\pi{\bm \Lambda} \subset \CH^2_\BC(M) \,\simeq\, \BC^{b_1}\,,
\end{equation}
where
\begin{equation}\label{LATTICE}
{\bf \Lambda} \,=\, \BL^\vee \,\oplus\, i \, e^2 \*\BL\,.
\end{equation}
Because the coupling $e^2$ appears in the definition of
the complex lattice ${\bm \Lambda}$, the physical interpretations of
the relations in \eqref{PERD} and \eqref{SHFTL} are very different.
The periodicity of $\Theta_M(\upgamma)$ under shifts in $\BL^\vee$ is a
classical property, visible already from the classical action \eqref{TOTAC} for
the scalar field.  Conversely, the quasi-periodicity of 
$\Theta_M(\upgamma)$ under shifts in ${i \, e^2 \*\BL}$ is a quantum
effect, which relies upon the sum over winding-sectors in the scalar
partition function.

The theta-function $\Theta_M(\upgamma)$ definitely depends on the
Riemannian structure on $M$, but this dependence occurs only through the
matrix of $L^2$ inner-products in \eqref{BIGSQ},
\begin{equation}
\SQ_{j k} \,=\, \left(\Fe_j, \Fe_k\right) \,=\, \int_M
\Fe_j\^\*\Fe_k\,,\qquad\qquad j,k = 1,\ldots,b_1\,,
\end{equation}
where we have selected an integral basis for ${\BL \simeq \BZ \Fe_1
  \oplus \cdots \oplus \BZ \Fe_{b_1}}$.  In terms of the basis for
$\BL$ and the ${b_1 \times b_1}$ matrix $\SQ$, we can write
$\Theta_M(\upgamma)$ very concretely as a sum over a vector
${\vec{n}\in\BZ^{b_1}}$ of integers,
\begin{equation}\label{GEOMTH}
\Theta_M(\upgamma) \,=\, \sum_{\vec{n}\in\BZ^{b_1}} \,
\exp{\!\left[-\pi \, e^2 \, \SQ_{j k} \, n^j \, n^k \,+\, i \,
    \upgamma_j \, n^j\right]}\,.
\end{equation}
In this expression, ${\upgamma_j\in\BC}$ for ${j = 1,\ldots,b_1}$ are the
components of the complex two-form $\upgamma$, expressed dually with
respect to the basis for $\BL$, 
\begin{equation}
\upgamma_j \,=\, \langle\upgamma, \Fe_j\rangle \,=\, \int_M
\upgamma\^\Fe_j\,.
\end{equation}

For instance, if $M$ has the rational homology of ${S^1 \times S^2}$,
then $\Theta_M(\gamma)$ reduces to the classical Jacobi theta-function
\begin{equation}\label{JACOBI}
\Theta(z;\tau) \,=\, \sum_{n=-\infty}^{+\infty} \, \exp{\!\left[\pi 
    i \, n^2 \, \tau + 2\pi i \, n \, z\right]}\,,
\end{equation}
evaluated at the purely-imaginary complex structure
\begin{equation}
\tau \,=\, i \, \frac{e^2 \ell^3}{R^2} \,\in\, i\,\BR\,,\qquad\qquad z \,=\,
\frac{\upgamma}{2\pi} \,\in\,\BC\,.
\end{equation}
Here $R$ is a length scale naturally identified with the radius of
$S^1$ when ${M = S^1 \times S^2}$.  More generally, if we introduce
the standard multi-variable extension of \eqref{JACOBI},
\begin{equation}\label{JACOBII}
\Theta(\vec{z};\Omega) \,=\, \sum_{\vec{n}\in\BZ^{b_1}}
\exp{\!\left[\pi \, i \, \Omega_{j k} \, n^j \, n^k \,+\, 2 \pi i \,
    z_j \, n^j\right]}\,,\qquad\qquad \vec{z} \,\in\,\BC^{b_1}\,,
\end{equation}
where $\Omega$ is a complex matrix with positive-definite
imaginary part, then the geometric theta-function $\Theta_M(\upgamma)$
in \eqref{GEOMTH} agrees with the classical theta-function
$\Theta(\vec{z};\Omega)$ under the assignments 
\begin{equation}\label{PERIDOM}
\Omega \,=\, i \, e^2 \, \SQ\,,\qquad\qquad \vec{z} \,=\,
\frac{\upgamma}{2\pi}\,.
\end{equation}

\medskip\noindent{\sl The Modular Transform of
  $\Theta_M(\upgamma)$}\medskip

Since $\Theta_M(\upgamma)$ agrees with the classical Jacobi
theta-function  $\Theta(\vec{z};\Omega)$ when the period matrix $\Omega$ is
imaginary, $\Theta_M(\upgamma)$ also inherits the well-known modular
properties of $\Theta(\vec{z};\Omega)$.  A concise exposition of the latter
can be found in Ch.\,$2.5$ of \cite{Mumford:83}, whose notation I follow.

Not surprising for our discussion of abelian duality,
the most important modular property will be the transformation of
$\Theta_M(\upgamma)$ under the analogue of the S-duality ${\tau
  \mapsto -1/\tau}$, acting here on the period matrix $\Omega$ by 
\begin{equation}\label{SOMEG}
S:\Omega \,\longmapsto\, -\Omega^{-1}\,.
\end{equation}
If $\Omega$ is purely imaginary as in \eqref{PERIDOM}, then this
feature is preserved under \eqref{SOMEG}, so that S-duality also acts
on the geometric theta-function $\Theta_M(\upgamma)$ by the inversion
\begin{equation}\label{SOMEQ}
S:\SQ \,\longmapsto\, e^{-4} \, \SQ^{-1}\,.
\end{equation}
At this stage, one could simply refer to the literature on
theta-functions to determine the transformation of
$\Theta_M(\upgamma)$ under the operation in 
\eqref{SOMEQ}.  However, for the convenience of the reader, I shall 
provide a brief derivation of the required transformation law.  

As well-known, the transformation of the theta-function under the inversion
in \eqref{SOMEG} or \eqref{SOMEQ} can be understood as a consequence of Poisson
resummation, which itself follows from the distributional identity
\begin{equation}
\sum_{n\in\BZ} \delta(x - n) \,=\, \sum_{m \in \BZ} \e{2\pi i m x}\,.
\end{equation}
Applied to the concrete description of $\Theta_M(\upgamma)$ in \eqref{GEOMTH},
this identity implies 
\begin{equation}\label{MODTII}
\begin{aligned}
\Theta_M(\upgamma) \,&=\, \sum_{\vec{n}\in\BZ^{b_1}} \,
\exp{\!\left[-\pi \, e^2 \, \SQ_{j k} \, n^j \, n^k \,+\, i \,
    \upgamma_j \, n^j\right]}\,,\\
&=\,  \sum_{\vec{n}\in\BZ^{b_1}} \, \int_{\BR^{b_1}} \!\!d^{b_1}\!x \,\, \delta(\vec{x} - \vec{n}) \, \exp{\!\left[-\pi \, e^2 \, \SQ_{j k} \, x^j \, x^k \,+\, i \,
    \upgamma_j \, x^j\right]}\,,\\
&=\, \sum_{\vec{m}\in\BZ^{b_1}} \, \int_{\BR^{b_1}}\!\!d^{b_1}\!x \,
\exp{\!\left[-\pi \, e^2 \, \SQ_{j k} \, x^j \, x^k \,+\, 2\pi i
    \left(m_j \,+\, \frac{\upgamma_j}{2\pi}\right) x^j \right]}\,.
\end{aligned}
\end{equation}
We evaluate the Gaussian integral over ${\vec{x}\in\BR^{b_1}}$ in the
last line of \eqref{MODTII} to obtain 
\begin{equation}\label{sDUALT}
\Theta_M(\upgamma) \,=\, \sum_{\vec{m}\in\BZ^{b_1}}
\frac{1}{e^{b_1}\sqrt{\det\SQ}} \,
\exp{\!\left[-\frac{\pi}{e^2}\left(\SQ^{-1}\right){}^{\!j k}\left(m_j
      \,+\, \frac{\upgamma_j}{2\pi}\right)\left(m_k \,+\,
      \frac{\upgamma_k}{2\pi}\right)\right]}\,.
\end{equation}

A more geometric interpretation for the right-hand side of \eqref{sDUALT}
follows once we recognize $\SQ^{-1}$ as the matrix whose
elements encode the $L^2$ inner-products of the basis for $\BL^\vee$
which is dual to the chosen basis for $\BL$,
\begin{equation}\label{DUALB}
\BL^\vee \,\simeq\, \BZ{\Fe}^{* 1} \oplus \cdots \oplus
\BZ{\Fe}^{* b_1}\,,\qquad\qquad \big\langle{\Fe}^{* j},
\Fe_k\big\rangle 
\,=\, \delta^j_k\,,
\end{equation}
so that
\begin{equation}\label{QINVS}
\left(\SQ^{-1}\right){}^{\!j k} \,=\, \left({\Fe}^{*j}, {\Fe}^{* k}\right) \,=\,
  \int_M {\Fe}^{* j} \^\*{\Fe}^{* k}\,,\qquad\qquad j,\,k \,=\, 1,\ldots,b_1\,.
\end{equation}
Here  in \eqref{DUALB} we introduce the Kronecker-delta, and we observe
that integrality for the dual basis
$\left\{{\Fe}^{* 1},\cdots,{\Fe}^{* b_1}\right\}$ of $\BL^\vee$ is a 
non-trivial consequence of Poincar\'e duality on $M$.  Otherwise, the
interpretation for $\SQ^{-1}$ in \eqref{QINVS} derives from the
tautological relation
\begin{equation}
\Fe^{* j} \,=\, \left(\SQ^{-1}\right){}^{\! j
  k}\left(\Fe_k,\,\cdot\,\right) \,\in\, \BL^\vee\,.
\end{equation}

As a result of \eqref{sDUALT} and \eqref{QINVS}, the geometric theta-function
$\Theta_M(\upgamma)$ on $M$ can be written not only in terms of a sum
over the lattice $\BL$, but also in terms of a sum over the dual
lattice $\BL^\vee$,
\begin{equation}\label{DLTHI}
\Theta_M(\upgamma) \,=\, \frac{1}{e^{b_1}\sqrt{\det\SQ}} \, \sum_{\lambda
  \in \BL^\vee} \exp{\!\left[-\frac{\pi}{e^2}\left(\lambda \,+\,
      \frac{\upgamma}{2\pi},\,\lambda \,+\,
      \frac{\upgamma}{2\pi}\right)\right]}\,,
\end{equation}
where $\left(\,\cdot\,,\,\cdot\,\right)$ now indicates the $L^2$-norm
on ${\BL^\vee \subset \CH^2(M)}$.  

Of course, we also recall the description of the other theta-function
$\Theta^\vee_M(\upzeta)$ which enters the Maxwell partition function,
\begin{equation}\label{DLTHII}
\Theta^\vee_M(\upzeta) \,=\, \sum_{\lambda\in\BL^\vee}
\exp{\!\left[-\frac{\pi}{e^2}\left(\lambda,\lambda\right) +
    i\,\langle\upzeta,\lambda\rangle\right]}\,.
\end{equation}
Comparing \eqref{DLTHI} and \eqref{DLTHII}, we see that 
\begin{equation}\label{RELTHS}
\Theta_M^{}(\upgamma) \,=\,  \frac{1}{e^{b_1}\sqrt{\det\SQ}} \,
\exp{\!\left[-\frac{\left(\upgamma,\upgamma\right)}{4\pi e^2}\right]}
\cdot \Theta^\vee_M\!\left(\frac{i}{e^2}\,\*\upgamma\right).
\end{equation}
The identification ${\upzeta = i\,\*\upgamma/e^2}$ in the
argument of $\Theta_M^\vee$ agrees with our conventions for $\upalpha$
and $\upbeta$ in Sections \ref{Scalar} and \ref{Vector}.

\subsection{The Role of Torsion}

The modular relation between $\Theta_M^{}$ and $\Theta^\vee_M$ in
\eqref{RELTHS} is the fundamental result which we need to compare the
respective scalar and vector partition functions $Z_M^{}$ and $Z_M^\vee$ under
duality. With the identification of parameters 
\begin{equation}
\upzeta \,=\, \frac{i}{e^2}\,\*\upgamma\,,
\end{equation}
we compute the ratio 
\begin{equation}
\frac{Z^\vee_M\!\left(\upzeta\right)}{Z_M\!\left(\upgamma\right)}
\,=\, \frac{\Delta^\vee_M \cdot \Theta^\vee_M(\upzeta)}{\Delta_M
  \cdot \Theta_M(\upgamma)}\,=\,
\frac{\Delta^\vee_M}{\Delta_M} \cdot e^{b_1} \sqrt{\det\SQ} \cdot \exp{\!\left[\frac{\left(\upgamma,\upgamma\right)}{4\pi e^2}\right]}\,.
\end{equation}
From \eqref{DMFINII} and \eqref{FDELVM}, the ratio of the respective Gaussian
factors $\Delta_M^{}$ and $\Delta^\vee_M$ is
\begin{equation}
\frac{\Delta^\vee_M}{\Delta_M} \,=\, \Tor_M \cdot \frac{1}{e^{b_1} 
  \ell^3} \sqrt{\det\SQ} \cdot
\frac{\left[{\det}'(\triangle_0)\right]^{3/2}}{\left[{\det}'(\triangle_1)\right]^{1/2}}\,.
\end{equation}
Thus,
\begin{equation}\label{RATZS}
\frac{Z^\vee_M\!\left(\upzeta\right)}{Z_M\!\left(\upgamma\right)} \,=\, \Tor_M
\cdot \frac{\det\SQ}{\ell^3} \,
\frac{\left[{\det}'(\triangle_0)\right]^{3/2}}{\left[{\det}'(\triangle_1)\right]^{1/2}} \cdot \exp{\!\left[\frac{\left(\upgamma,\upgamma\right)}{4\pi
      e^2}\right]}\,.
\end{equation}
We now reach the most important question in the present paper.  

When is the ratio of partition functions in \eqref{RATZS} equal to one?  

If $Z_M^{}(\upgamma)$ is to be equal to $Z^\vee_M(\upzeta)$, then evidently
${\upgamma = \upzeta = 0}$ in \eqref{RATZS}.  In that case, the
expression on the right-hand side of \eqref{RATZS} does not depend of
the Maxwell coupling $e^2$, and the ratio reduces to the product  
\begin{equation}\label{RATZSII}
\frac{Z^\vee_M(0)}{Z_M(0)} \,=\, \Tor_M \cdot \tau_M\,,\qquad\qquad
\tau_M \,=\, \frac{\det\SQ}{\ell^3} \,
\frac{\left[{\det}'(\triangle_0)\right]^{3/2}}{\left[{\det}'(\triangle_1)\right]^{1/2}}\,. 
\end{equation}
As before, $\Tor_M$ is the number of elements in the torsion subgroup
$H^2(M;\BZ)_{\rm tors}$, an obvious topological invariant of $M$.  So the
remaining factor to examine is the mysterious ratio $\tau_M$ of
functional determinants in \eqref{RATZSII}

Manifestly, $\tau_M$ depends only upon the Riemannian metric on
$M$.  By construction, both $Z_M^{}$ and $Z^\vee_M$ are invariant under
the combined scalings in \eqref{SCALEg} and \eqref{SCALEe}.  Hence
$\tau_M$ must also be preserved  by the scale transformation ${g
  \mapsto \Lambda^2\,g}$ of the metric in \eqref{SCALEg}.  This feature suggests that
$\tau_M$, like the quantity $\Tor_M$ in \eqref{RATZSII}, could be a
topological invariant of $M$.  In fact, as I now explain, $\tau_M$ is
precisely the Reidemeister torsion of the three-manifold, evaluated on
an integral basis for the cohomology.

\medskip\noindent{\sl Reidemeister vs Ray-Singer Torsion}\medskip

Before discussing $\tau_M$, let me briefly recall a few facts about
Reidemeister torsion.  Nice expositions on Reidemeister torsion can be found in
\cite{Milnor:66,Freed:92,Nicolaescu:03}, and a prominent application
of these ideas to gauge theory on a Riemann surface appears in
\cite{Witten:1991we}.  Here I follow the presentation of Freed in
\cite{Freed:92}, specialized to dimension three for concreteness.

The Reidemeister torsion is a combinatorial invariant of $M$, defined
in terms of the chain complex $C_\bullet$ associated to a given triangulation (or
cellular structure) on $M$,
\begin{equation}\label{CHAIN}
C_\bullet:
0\,\longrightarrow\,C_3\buildrel\partial\over\longrightarrow\,C_2\,\buildrel\partial\over\longrightarrow\,C_1\,\buildrel\partial\over\longrightarrow\,
C_0 \,\longrightarrow\, 0\,.
\end{equation}
We work with real coefficients throughout, and the homology of this
chain complex is $H_*(M;\BR)$.

Each $C_j$ for ${j=0,\ldots,3}$ is a finite-dimensional vector
space with a distinguished set of generators, the elementary 
simplices in $M$.  Because the vector space $C_j$ has a
basis, $C_j$ also has an associated metric, for which the generating
simplices are orthonormal.  Accompanying the metric on $C_j$ is a volume form
${\nu_j\in (\det C_j)^{-1}}$.  Here ${\det C_j = \bigwedge^{\rm
    top}C_j}$ indicates the top exterior power.  As standard in this
business, we will not worry about orientations or signs; by
convention, the torsion will be positive.

I first describe the torsion assuming the complex $C_\bullet$ to
be acyclic, with trivial homology.  Let $k_j = \dim\Im\partial\!:C_j
  \to C_{j-1}$, and pick an element ${s_j \in \bigwedge^{k_j}C_j}$ so
  that ${\partial s_j \neq 0}$.  We now consider the following element in
  the alternating tensor product,
\begin{equation}
u \,=\, \bigotimes_{j=0}^3 \left(\partial s_{j+1}\^ s_j\right)^{(-1)^j}
\,\in\, \bigotimes_{j=0}^3 \left(\det C_j\right)^{(-1)^j}\,.
\end{equation}
The element $u$ is independent of the choices of the $s_j$, so we can
define the torsion of the acyclic complex $C_\bullet$ as 
\begin{equation}
\tau(C_\bullet) \,=\, u \otimes \bigotimes_{j=0}^3 \nu_j^{(-1)^j}\,\in\,\BR\,.
\end{equation}

Of course, in the geometric situation $C_\bullet$ always has
non-trivial homology, since ${H_0(M) 
  = H_3(M) = \BZ}$ for a closed, orientable three-manifold.  To define
$\tau(C_\bullet)$ more generally when $C_\bullet$ has homology, we
split $C_\bullet$ as ${C_\bullet = C'_\bullet \oplus C''_\bullet}$,
where $C'_\bullet$ is acyclic and the differential on $C''_\bullet$
vanishes, ${\partial\big|_{C''_\bullet} = 0}$.  Hence $C''_j$ is
isomorphic to $H_j(M;\BR)$.  We now apply the preceding construction to
the acyclic summand $C'_\bullet$, with 
$k_j = \dim\Im\partial\!:C'_j\to C'_{j-1}$, ${s_j \in \bigwedge^{k_j}
  C'_j}$ satisfying ${\partial s_j   \neq 0}$, and 
\begin{equation}\label{UTORRS}
u \,=\, \bigotimes_{j=0}^3 \left(\partial s_{j+1}\^s_j\right)^{(-1)^j}
\,\in\, \bigotimes_{j=0}^3 \left(\det C'_j\right)^{(-1)^j}\,.
\end{equation}
The torsion $\tau(C_\bullet)$ is then defined as the element 
\begin{equation}\label{RTORRS}
\tau(C_\bullet) \,=\, u \otimes \bigotimes_{j=0}^3 \nu_j^{(-1)^j}
\,\in\, \bigotimes_{j=0}^3 \left[\det H_j(M;\BR)\right]^{(-1)^{j+1}}\,.
\end{equation}
Equivalently, $\tau(C_\bullet)$ lies in the dual of the determinant line 
\begin{equation}
\det H_*(M;\BR) \,=\, \bigotimes_{j=0}^3 \left[\det H_j(M;\BR)\right]^{(-1)^j}\,.
\end{equation}
Finally, though we have defined the torsion for the chain complex
$C_\bullet$ associated to a particular triangulation of $M$, one
checks that \eqref{RTORRS} is invariant under any refinement of the
triangulation, and hence ${\tau_M = |\tau(C_\bullet)|}$ defines a
smooth invariant of $M$.

Let us be more explicit about where $\tau_M$ in \eqref{RTORRS} is
valued.  In dimension three, the dual of the determinant line is given
(with the obvious abbreviations) by 
\begin{equation}
\begin{aligned}
\left(\det H_*\right){}^{\!-1} &= \left[\det H_0 \otimes
  \left(\det H_1\right){}^{\!-1}\!\otimes \det H_2 \otimes \left(\det
  H_3\right){}^{\!-1}\right]{}^{\!-1},\\
&\simeq \left[\det H^0 \otimes \left(\det H^1\right){}^{\!-1}\!\otimes \det
H^2 \otimes \left(\det H^3\right){}^{\!-1}\right],\\
\end{aligned}
\end{equation}
where we apply the canonical duality between $H_*(M;\BR)$ and
$H^*(M;\BR)$ in the second step.  But Poincar\'e duality on $M$ also
implies the isomorphisms
\begin{equation}
\begin{aligned}
\det H^2(M;\BR) \,&\simeq\, \left[\det
  H^1(M;\BR)\right]{}^{\!-1}\,,\\
\det 
H^3(M;\BR) \,&\simeq\, \left[\det H^0(M;\BR)\right]{}^{\!-1}\,.
\end{aligned}
\end{equation}
As a result, $\tau_M$ is valued in the one-dimensional vector space
\begin{equation}\label{DELT}
\tau_M \,\in\, \left[\det H^0(M;\BR) \otimes \left(\det
  H^1(M;\BR)\right){}^{\!-1}\right]{}^{\!\otimes 2}\,.
\end{equation}
To assign a value to $\tau_M$ as a real number, we evaluate $\tau_M$
in \eqref{DELT} on an integral basis for $H^1(M;\BR)$ and
$H_0(M;\BR)$, the latter corresponding simply to the choice of a point
${p \in M}$.  By the standard properties of the determinant, the
result is independent of the choice of integral basis.

The description thus far of $\tau_M$ is combinatorial, depending upon
the choice of a triangulation for $M$.  This description is most
useful for computations in examples.  On the other hand, the quantity which
actually appears in \eqref{RATZSII} is analytic in character, 
\begin{equation}\label{TAUM}
\tau_M \,=\, \frac{\det\SQ}{\ell^3} \,
\frac{\left[{\det}'(\triangle_0)\right]^{3/2}}{\left[{\det}'(\triangle_1)\right]^{1/2}}\,.
\end{equation}
Most famously, the ratio of functional
determinants appearing in \eqref{TAUM} is the Ray-Singer analytic\footnote{Some
  authors define the analytic torsion as the logarithm of $T_M$, but I follow the convention
  already established for the Reidemeister torsion.} torsion
\cite{Ray:70,Ray:71} 
\begin{equation}
T_M \,=\,
\frac{\left[{\det}'(\triangle_0)\right]^{3/2}}{\left[{\det}'(\triangle_1)\right]^{1/2}}\,,
\end{equation}
and the equality between the combinatorial \eqref{RTORRS} and analytic
\eqref{TAUM} descriptions of $\tau_M$ is a consequence of the
Cheeger-M\"uller theorem \cite{Cheeger:77,Cheeger:79,Muller:78}
relating Reidemeister to Ray-Singer torsion.

As the reader may note, the Reidemeister torsion $\tau_M$ and the
Ray-Singer torsion $T_M$ are not precisely equal in our situation, but
instead obey 
\begin{equation}\label{CORRCT}
\tau_M \,=\, \frac{\det\SQ}{\ell^3} \cdot T_M\,.
\end{equation}
The correction factor $\det\SQ/\ell^3$ is discussed
in Appendix B of \cite{Friedmann:2002ty} and arises due to the
non-trivial homology of $M$.  Very briefly, both $\tau_M$ and $T_M$
are intrinsically valued in the dual of the determinant line $\det
H_*(M;\BR)$, identified concretely in \eqref{DELT}.  To assign real
values to $\tau_M$ and $T_M$, we must pick a basis for the cohomology
on which we evaluate the torsions.  For $\tau_M$ we naturally use an
integral basis, and for $T_M$ we use a basis which is orthonormal with
respect to the $L^2$ inner-product.  The correction factor in
\eqref{CORRCT} is necessary to relate these different choices of basis.

Explicitly, let $\SA_0$ and $\SA_1$ be linear maps which express integral
bases for $H^0(M;\BR)$ and $H^1(M;\BR)$ in terms of $L^2$-bases for
the same spaces.  Evaluating $\tau_M$ and $T_M$ on the respective
bases, we see that $\tau_M$ and $T_M$ satisfy
\begin{equation}\label{ATTS}
\tau_M \,=\, \left(\frac{\det \SA_1}{\det \SA_0}\right)^{\!2} T_M\,.
\end{equation}
The square in \eqref{ATTS} appears due to the corresponding
square in the determinant line in \eqref{DELT}.  

The constant $\SA_0$ can be evaluated directly.  The integral
generator for $H^0(M;\BR)$ is the constant function $1$, and the
$L^2$-generator for $H^0(M;\BR)$ is the constant
function $\ell^{-3/2}$ (where $\ell^3$ is the volume of $M$), so 
\begin{equation}\label{BIGAZ}
\SA_0 \,=\, \ell^{3/2}\,.
\end{equation}
As for $\SA_1$, we have already introduced integral generators
$\{\Fe_1,\ldots,\Fe_{b_1}\}$ for $H^1(M;\BR)$ in \eqref{INTB}.  If
$\{\omega_1,\ldots,\omega_{b_1}\}$ is a basis for
$H^1(M;\BR)$ which is orthonormal with respect to the $L^2$
inner-product, then by definition 
\begin{equation}
\Fe_j \,=\, \left(\SA_1\right){}^{\! j'}_j \, \omega_{j'}^{}\,,\qquad\qquad j,
j'\,=\,
1\,,\ldots\,, b_1\,,
\end{equation}
and 
\begin{equation}\label{BIGAO}
\SQ_{j k} \,=\, \left(\Fe_j,\Fe_k\right) \,=\,
\left(\SA_1^t\SA_1^{}\right){}_{\!j k}\,.
\end{equation}
Here $\SA^t_1$ is the transpose of $\SA_1^{}$.  Together, the
relations in \eqref{ATTS}, \eqref{BIGAZ}, and \eqref{BIGAO} produce
the metric-dependent correction factor in \eqref{CORRCT}.

\medskip\noindent{\sl Duality for the Partition Function}\medskip

To summarize, the ratio of the Maxwell to scalar partition function
on $M$ is a topological invariant,
\begin{equation}
\frac{Z^\vee_M(0)}{Z_M(0)} \,=\, \Tor_M \cdot \tau_M\,,\qquad\qquad
\Tor_M \,=\, \big|H^2(M;\BZ)_{\rm tors}\big|\,,
\end{equation}
where $\Tor_M$ is the number of elements in $H^2(M;\BZ)_{\rm tors}$,
and $\tau_M$ is the Reidemeister torsion evaluated with respect to an integral
basis for the cohomology of $M$.  Via its combinatorial definition,
the Reidemeister torsion is eminently computable, and I claim 
\begin{equation}\label{FORMTA}
\tau_M \,=\, \frac{1}{\Tor_M}\,.
\end{equation}
Compare also to Theorem $2.39$ in \cite{Nicolaescu:03} and the
surrounding discussion of the Alexander formula.\footnote{I thank
  Maxim Braverman for pointing out this formula to me.}  As a result, abelian
duality for the partition function holds globally in the most naive
sense, with 
\begin{equation}
Z^\vee_M(0) = Z_M^{}(0)\,.
\end{equation}
This result strengthens the conclusions in \cite{BrodaWG} beyond the
case ${b_1=0}$.

When $M$ is a rational homology sphere, with ${b_1 = 0}$, a short
proof of the formula \eqref{FORMTA} for $\tau_M$ appears in
\cite{Witten:2003ya}, but the proof can be easily generalized to the
arbitrary closed, orientable three-manifold.  Such a manifold can
always be given a cellular structure with exactly one 0-cell and one
3-cell, so that the chain complex $C_\bullet$ in \eqref{CHAIN} takes
the concrete form
\begin{equation}\label{CHAINTW}
C_\bullet:
0\,\longrightarrow\,\BZ\,\buildrel
0\over\longrightarrow\,\BZ^N\,\buildrel \partial\over\longrightarrow\,\BZ^N\,\buildrel
0\over\longrightarrow\,\BZ\,\longrightarrow\, 0\,.
\end{equation}
Geometrically, a chain complex of the form in \eqref{CHAINTW} is
naturally associated to any Heegaard decomposition of ${M = H_1 \cup H_2}$ into
handlebodies $H_1$ and $H_2$.  Because ${H_0(M) = H_3(M) = \BZ}$, the
indicated maps in \eqref{CHAINTW} vanish, and
Poincar\'e duality otherwise implies that the chain groups satisfy ${C_1,
  C_2 \simeq \BZ^N}$ for some $N$.  Once we select integral generators
$\{v_1,\cdots,v_N\}$ and $\{w_1,\cdots,w_N\}$ for $C_1$ and $C_2$ to
realize the isomorphism with $\BZ^N$, the
non-trivial differential $\partial$ in \eqref{CHAINTW} can 
be identified concretely with a square, ${N \times N}$ integer
matrix.

We first consider the simpler case that ${b_1 = 0}$.  Necessarily,
$\partial$ is injective and has full rank over $\BR$.  The first
homology group ${H_1(M) \simeq C_1/\Im(\partial)}$ is purely torsion, 
and by the Universal Coefficient Theorem, ${\Tor_M =
  \big|H_1(M)\big|}$.  Via standard arguments, the number of
elements in the lattice quotient $C_1/\Im(\partial)$ is the volume of
the simplex spanned by the images
$\left\{\partial(w_1),\cdots,\partial(w_N)\right\}$ of the generators
for $C_2$ under $\partial$.  Concretely, this volume can be computed
as the absolute value of the determinant of $\partial$ as an 
${N\times N}$ matrix, 
\begin{equation}
\Tor_M = \left|\det\partial\right|\,,\qquad\qquad b_1 = 0\,.
\end{equation}

By comparison, to evaluate the Reidemeister torsion, we pick any point
${p\in M}$ to generate $H_0(M)$, and we use $M$ itself to generate
$H_3(M)$.  Because we have only one 0-cell and one 3-cell, both of
which represent the homology, $C_0$ and $C_3$ both contribute `$1$' to the formula for $\tau_M$ in \eqref{RTORRS}.  To treat
$C_1$ and $C_2$, we trivialize the determinant lines $\det
C_1, \det C_2 \simeq \BR$ with the volume forms $\nu_1 = v_1\^\cdots\^v_N$ and
$\nu_2 = w_1\^\cdots\^w_N$.  For any choice of ${s_2 \in \det C_2}$,
the formula for $\tau_M$ in \eqref{UTORRS} and \eqref{RTORRS} then
reduces to the ratio 
\begin{equation}
\tau_M \,=\, \left|\frac{s_2}{\partial s_2}\right| \,=\,
\frac{1}{\left|\det\partial\right|}\,,\qquad\qquad s_2 \,\in\,\det{C_2}\,.
\end{equation}
Hence ${\tau_M = 1/\Tor_M}$ if ${b_1 = 0}$.

When $b_1$ is non-vanishing, the ideas leading to \eqref{FORMTA} are
much the same, except for the fact that the determinant of $\partial$
now vanishes.  By assumption, ${\partial:C_2\to C_1}$ has a non-trivial kernel,
with ${H_2(M) = \ker\partial}$.  Clearly $H_2(M) \simeq \BZ^{b_1}$ is 
generated freely, and we assume without loss that the subset
$\left\{w_1,\cdots,w_{b_1}\right\}$ generates the kernel of $\partial$
in $C_2$.  Hence $\left\{w_1,\cdots,w_{b_1}\right\}$ represent
integral generators for $H_2(M)$.  On the other hand, $H_1(M) =
C_1/\Im\partial$ may still contain a torsion subgroup ${H_1(M)_{\rm
    tors} \,\simeq\, H^2(M;\BZ)_{\rm tors}}$, with ${H_1(M;\BR) \simeq
  \BR^{b_1}}$.  Again without loss, we assume that the generating
subset $\{v_1,\cdots,v_{b_1}\}$ spans the cokernel of $\partial$ over
$\BR$.  Thus $\{v_1,\cdots,v_{b_1}\}$ represent integral generators for $H_1(M)$
modulo torsion.  

As before, the number of elements in $H_1(M)_{\rm
  tors}$ can be interpreted geometrically as the volume of the $\left(N -
  b_1\right)$-dimensional simplex in $\BR^N$ spanned by the image of $\partial$
on the integral generators for $C_2$.  To compute that volume, we
extend $\partial$ linearly to a non-degenerate integral map
$\widehat\partial:C_2\to C_1$ via the assignments 
\begin{equation}
\widehat\partial(w_1) = v_1\,,\qquad \widehat\partial(w_2) =
v_2\,, \qquad\ldots\,, \qquad\widehat\partial(w_{b_1}) = v_{b_1}\,,
\end{equation}
as well as ${\widehat\partial = \partial}$ for all other generators of
$C_2$.  By construction, the determinant of $\widehat\partial$ is
non-vanishing and computes the requisite volume,
\begin{equation}\label{BIGT}
\Tor_M \,=\, \big|\!\det\widehat\partial\big|,\qquad\qquad b_1 \neq 0\,.
\end{equation}

With our choices for homology generators, the description for $\tau_M$
in \eqref{UTORRS} and \eqref{RTORRS} similarly unravels to the ratio
\begin{equation}\label{BIGTAU}
\tau_M \,=\, \left|\frac{s_2 \^ w_1 \^ \cdots \^ w_{b_1}}{\partial s_2 \^
  v_1 \^ \cdots \^ v_{b_1}}\right| \,=\,
\frac{1}{\big|\!\det\widehat\partial\big|}\,,\qquad\qquad s_2 \,\in\,
\det C_2'\,.
\end{equation}
Here $C_2'$ is the $(N-b_1)$-dimensional subspace of $C_2$ spanned by
all generators other than $\left\{w_1,\cdots,w_{b_1}\right\}$, and
$s_2$ is any non-vanishing element in the determinant line $\det
C_2'$.  Comparing \eqref{BIGT} and \eqref{BIGTAU}, we obtain the
promised reciprocal relation between the torsion
invariants $\Tor_M$ and $\tau_M$.

\medskip
\section{Path Integral Explanation}\label{Path}

By somewhat laborious direct calculations, we have obtained the dual
identity
\begin{equation}\label{DUALZS}
Z^\vee_M(\upzeta) \,=\, Z^{}_M(\upgamma) \cdot
\exp{\!\left[\frac{\left(\upgamma,\upgamma\right)}{4\pi
      e^2}\right]}\,,\qquad\qquad \upzeta \,=\, \frac{i}{e^2}\,\*\upgamma\,.
\end{equation}
The relation between the scalar and the Maxwell partition functions
can also be derived more economically by formal path integral
manipulations of the sort in \cite{RocekPS,WittenGF}.  This approach
to abelian duality in three dimensions appears already in Lecture 8 of
\cite{Deligne}, but I take the opportunity now to review it.

As one application, the path integral perspective on duality neatly
explains the otherwise anomalous exponential factor involving
$\upgamma$ in \eqref{DUALZS}, which arose from the modular
transformation of $\Theta_M(\upgamma)$ in Section \ref{Modular}.
In preparation for the Hamiltonian analysis in \cite{BeasleyII}, I
conclude Section \ref{Path} with a discussion of duality for three
natural classes of operators on $M$.

\subsection{Duality for the Partition Function}\label{DUALPIP}

We start with the path integral which describes the scalar partition
function on $M$,
\begin{equation}
Z_M(\upgamma) \,=\,
\sum_{\omega\in\BL}\,\int_{\CX_\omega}\!\!\CD\phi\,\exp{\!\big[-{\bf
    I}_{\rm tot}(\phi)\big]}\,,
\end{equation}
where the sigma model action for ${\phi:M \to S^1}$ is given by 
\begin{equation}\label{INOUGHT}
{\bf I}_{\rm tot}(\phi) \,=\, \frac{e^2}{4\pi}\left(d\phi,d\phi\right)
\,+\, \frac{1}{2\pi i}
\left\langle\upgamma,d\phi\right\rangle\,,\qquad \upgamma \,\in\,\CH^2_\BC(M)\,.
\end{equation}

To find a dual reformulation for the path integral, we enlarge the space
of fields to include a connection $B$ on the trivial $U(1)$-bundle $P_0$
over $M$.  Under a homotopically-trivial gauge transformation,
parametrized by a function ${f\in\Omega^0(M)}$, the pair
$\left(\phi,B\right)$ transforms by 
\begin{equation}\label{GAUGEB}
\phi \,\mapsto\, \phi \,+\, f\,,\qquad\qquad B\,\mapsto\, B \,-\, df\,.
\end{equation}
As a result, the combination ${d_B \phi = d\phi + B}$ is gauge-invariant.
For future reference, $\CG_B$ will denote the group of gauge
transformations acting on the pair $\left(\phi,B\right) \in
  \CX_\omega \times \CA(P_0)$.  Finally, the sigma model action in
\eqref{INOUGHT} can be promoted to a gauge-invariant action for the pair
$\left(\phi,B\right)$ by substituting the covariant derivative
$d_B\phi$ everywhere for $d\phi$,
\begin{equation}\label{IB}
{\bf I}_{\rm tot}\!\left(\phi,B\right) \,=\,
\frac{e^2}{4\pi}\left(d_B\phi,d_B\phi\right) 
\,+\, \frac{1}{2\pi i}
\left\langle\upgamma,d_B\phi\right\rangle\,,\qquad\qquad d_B\phi \,=\,
d\phi \,+\, B\,.
\end{equation}

The gauged sigma model for the pair $\left(\phi,B\right)$ with the
classical action in \eqref{IB} cannot be the whole story, because we
must also incorporate the Maxwell gauge field $A$ somehow.  So we let
$P_\lambda$ be the $U(1)$-bundle with Chern class ${\lambda \in
  H^2(M;\BZ)}$, and we let $A$ be a connection on $P_\lambda$.  To couple
$A$ to the pair $\left(\phi,B\right)$, we consider the mixed
Chern-Simons interaction 
\begin{equation}\label{CSAB}
{\SC\SS}(A,B) \,=\, \frac{1}{2\pi}\int_M F_A \^ B\,,\qquad\qquad F_A \,=\, dA\,.
\end{equation} 
Manifestly, ${\SC\SS}(A,B)$ is invariant under the group $\CG_A$ of
gauge transformations acting on $A$.  In addition, ${\SC\SS}(A,B)$ is
invariant under homotopically-trivial elements of $\CG_B$, and
otherwise the value of ${\SC\SS}(A,B)$ shifts by integral multiples
of $2\pi$ under ``large'' gauge transformations in $\CG_B$.  Thus
${\SC\SS(A,B)}$ is naturally valued in the circle,
\begin{equation}
\SC\SS(A,B) \,\in\, \BR/2\pi\BZ\,.
\end{equation}

The latter property is perhaps most transparent when ${\SC\SS}(A,B)$ is
considered via bordism.  Let $X$ be a four-manifold bounding $M$, over
which $A$ and $B$ extend.  Then alternatively,
\begin{equation}\label{BORDISM}
{\SC\SS}(A,B) \,=\, \frac{1}{2\pi} \int_X F_A \^ F_B\,,\qquad\qquad M
\,=\, \partial X\,.
\end{equation}
Integrality of both ${F_A/2\pi}$ and ${F_B/2\pi}$ ensures that the
intersection pairing in \eqref{BORDISM} is well-defined modulo $2\pi$,
regardless of the choice of $X$.  We also see that the mixed
Chern-Simons interaction in \eqref{CSAB} occurs at level one, the minimum 
for gauge-invariance in the absence of additional geometric
structure (eg.~a spin structure) on $M$.

 We now couple the connection $A$ to the pair $\left(\phi, B\right)$
 through the classical action 
\begin{equation}\label{INOAB}
{\bf I}_{\rm tot}\!\left(\phi,A,B\right) \,=\,
\frac{e^2}{4\pi}\left(d_B\phi,d_B\phi\right) 
\,+\, \frac{1}{2\pi i}
\left\langle\upgamma,d_B\phi\right\rangle \,-\, i\,{\SC\SS}(A,B)\,.
\end{equation}
By construction, the exponential of the classical action in \eqref{INOAB} is 
invariant under the product group ${\CG_A \times \CG_B}$, acting by
separate gauge transformations on $A$ and on the pair
$\left(\phi,B\right)$.  Associated to our classical action for $\left(\phi,
A, B\right)$ is the generalized partition function 
\begin{equation}\label{GENZP}
\begin{aligned}
\wt{Z}_M\!\left(\upgamma\right) &=\mskip -20mu
\sum_{\left(\omega,\lambda\right) \in \BL\oplus H^2(M;\BZ)} 
\frac{1}{\Vol\!\left(\CG_A\right)}\frac{1}{\Vol\!\left(\CG_B\right)}\int_{\CX_\omega\times
  \CA(P_\lambda) \times \CA(P_0)}\mskip -40mu\CD\phi\,\CD A\,\CD B \,
\exp{\!\left[-{\bf I}_{\rm tot}\!\left(\phi,A,B\right)\right]}\,.
\end{aligned}
\end{equation}
As indicated, the generalized partition function now involves a sum over all
winding sectors for $\phi$ as well as a sum over all topological types
for the $U(1)$-bundle on which $A$ is a connection.  We do not sum over
the topological type of the bundle for the auxiliary connection $B$,
though, for the following reason.  Since $\phi$ is assumed to be 
defined everywhere on $M$, the exponential $\e{i\phi}$ provides a global
section of the $U(1)$-bundle $P_0$ on which $B$ is a
connection.  This section trivializes $P_0$, which must therefore be a trivial
bundle.\footnote{I thank Edward Witten for this remark.}

The generalized partition function in \eqref{GENZP} can be studied in
two ways.

First, since the connection $A$ enters the classical action in \eqref{INOAB}
linearly through the Chern-Simons coupling $\SC\SS(A,B)$, the path
integral over $A$ can be performed directly.  As argued carefully in
\cite{Witten:2003ya}, the result is simply a delta-function that sets
$B$ to zero modulo gauge-equivalence,
\begin{equation}\label{DELBID}
\delta([B]) \,=\, \sum_{\lambda\in
  H^2(M;\BZ)} \frac{1}{\Vol\!\left(\CG_A\right)}\int_{\CA(P_\lambda)}\!\!\CD A \,
\exp{\!\big[i\,\SC\SS(A,B)\big]}\,.
\end{equation}
To explain this identity, we again decompose the arbitrary
connection $A$ on the bundle $P_\lambda$ as a sum 
\begin{equation}\label{HATA}
A \,=\, \widehat{A}_\lambda \,+\, \eta\,,
\end{equation}
where $\widehat{A}_\lambda$ is a fiducial connection with harmonic
curvature $2\pi\lambda$, and $\eta$ (like $B$) is a connection on the trivial
bundle $P_0$.  After we substitute for $A$ in \eqref{HATA} and
integrate by parts, the Chern-Simons pairing becomes 
\begin{equation}
\SC\SS(A, B) \,=\, \int_M \lambda\^ B \,+\, \frac{1}{2\pi}\int_M
\eta\^F_B\,,\qquad\qquad F_B = dB\,.
\end{equation}
Up to normalization, the path integral over $\eta$ produces a formal
delta-function that sets ${F_B = 0}$.  The remaining sum over
${\lambda \in H^2(M;\BZ)}$ produces a second delta-function
that requires $B$ to have trivial holonomy on $M$.  Hence $B$ is
gauge-equivalent to zero.  With a bit more work, one
can verify \cite{Witten:2003ya} that the coefficient of the
delta-function in \eqref{DELBID} is precisely one, but I omit
those details here.

After applying the identity in \eqref{DELBID} to the generalized
partition function in \eqref{GENZP}, we find
\begin{equation}\label{GENZPII}
\wt{Z}_M\!\left(\upgamma\right) \,=\,
\sum_{\omega \in\BL} \,\frac{1}{\Vol\!\left(\CG_B\right)} \int_{\CX_\omega \times \CA(P_0)}\mskip -15mu
\CD\phi\,\,\CD B\,\, \delta\!\left([B]\right) \, \exp{\!\left[-{\bf I}_{\rm
    tot}(\phi,B)\right]}\,.
\end{equation}
Because $B$ must be gauge-trivial due to the delta-function in the integrand of \eqref{GENZPII},
we can set ${B = 0}$ by an appropriate gauge transformation.
The path integral over $B$ then contributes $\Vol\!\left(\CG_B\right)$ to
cancel the prefactor in \eqref{GENZPII}, and we obtain the simple result 
\begin{equation}\label{TWOZS}
\wt{Z}_M(\upgamma)=\sum_{\omega \in H^1(M;\BZ)}\int_{\CX_\omega}\!\!\CD\phi\,
\exp{\!\left[-{\bf I}_{\rm tot}(\phi)\right]} \,=\, Z_M(\upgamma)\,.
\end{equation}
Thus $\wt{Z}_M(\upgamma)$ agrees with the scalar partition function
from Section \ref{Scalar}.

Alternatively, we return to the generalized partition function
\eqref{GENZP} and perform the respective path integrals over $\phi$
and $B$ instead.  Since $\phi$ is automatically gauge-trivial with
respect to the action of $\CG_B$, we set ${\phi =
  0}$ by a gauge transformation and cancel the 
prefactor ${1/\Vol\!\left(\CG_B\right)}$ to obtain
\begin{equation}\label{GENZPTHR}
\wt{Z}_M(\upgamma) \,=\, \sum_{\lambda\in H^2(M;\BZ)}
\frac{1}{\Vol\!\left(\CG_A\right)} \,\int_{\CA(P_\lambda) \times
  \CA(P_0)} \!\!\CD A\,\CD B\,\,\exp{\!\left[-{\bf I}_{\rm tot}(A,B)\right]}\,,
\end{equation}
where
\begin{equation}\label{TOTAB}
{\bf I}_{\rm tot}(A,B) \,=\, \frac{e^2}{4\pi}\left(B,B\right) \,+\,
\frac{1}{2\pi i} \left\langle\upgamma,B\right\rangle \,+\,
\frac{1}{2\pi i} \left\langle F_A,B \right\rangle
\end{equation}

The path integral over the auxiliary connection $B$ in
\eqref{GENZPTHR} and \eqref{TOTAB} is yet another Gaussian integral,
of a much simpler 
form than the Gaussian integrals which we analyzed in
Sections \ref{Scalar} and \ref{Vector}.  We immediately perform that 
integral to obtain a reformulation of $\wt Z_M(\upgamma)$ 
involving only the Maxwell gauge field,
\begin{equation}
\wt Z_M(\upgamma) \,=\, \sum_{\lambda\in H^2(M;\BZ)}
\frac{1}{\Vol\!\left(\CG_A\right)} \int_{\CA(P_\lambda)} \!\!\CD
A\,\,\exp{\!\left[-{\bf I}_{\rm tot}(A)\right]}\,.
\end{equation} 
To determine the action for $A$, we substitute the classical value for
${B = (i/e^2) (F_A + \upgamma)}$ into \eqref{TOTAB},
\begin{equation}\label{EFFA}
\begin{aligned}
{\bf I}_{\rm tot}(A) \,&=\, \frac{1}{4\pi e^2} \left(F_A \,+\,
  \upgamma,\, F_A \,+\, \upgamma\right),\qquad\qquad \upzeta \,=\,
\frac{i}{e^2}\*\upgamma\,,\\
&=\, \frac{1}{4\pi e^2} \left(F_A, F_A\right) \,+\, \frac{1}{2\pi
  i}\left\langle\upzeta,F_A\right\rangle \,+\, \frac{1}{4\pi
  e^2}\left(\upgamma, \upgamma\right).
\end{aligned}
\end{equation}
Comparing the classical action for $A$ in \eqref{EFFA} to the
corresponding action \eqref{TOTMC} from Section \ref{Vector}, we
deduce 
\begin{equation}
\wt{Z}_M^{}(\upgamma) \,=\,
Z_M^\vee(\upzeta)  \cdot \exp{\!\left[-\frac{1}{4\pi
      e^2}(\upgamma,\upgamma)\right]}\,,\qquad  \upzeta \,=\,
\frac{i}{e^2}\*\upgamma\,.
\end{equation}
Since $\wt{Z}_M^{}$ is equal to $Z_M^{}$, this relation reproduces
\eqref{DUALZS}.

\subsection{Duality for Operators}

To conclude, let us review the dual descriptions for three natural
classes of operators on the three-manifold $M$.  For simplicity in the 
following, I set the cohomological parameters $\upgamma$ and $\upzeta$ to zero.

\medskip\noindent{\sl Some Local and Non-Local Operators}\medskip

Of the three operators that we consider, two are
well-known:~the vertex operator and the Wilson loop operator.  The
vertex operator is the local operator described classically in the
sigma model by 
\begin{equation}\label{VERTEXP}
\SV_k(p) \,=\, \e{\! i k\phi(p)}\,,\qquad\qquad  k \,\in\,\BZ\,,
\end{equation}
for some point ${p \in M}$.  The condition that $\SV_k(p)$ be
single-valued under the shift ${\phi \mapsto \phi + 2\pi}$ requires
the parameter $k$ to be an integer.  Physically, $k$ labels the charge
of $\SV_k(p)$ under the global $U(1)$ symmetry which acts additively
on $\phi$ by a constant shift, 
\begin{equation}\label{GLBLU}
U(1): \phi\,\longmapsto\, \phi \,+\, c\,,\qquad\qquad
c\,\in\,\BR/2\pi\BZ\,.
\end{equation}

In Section \ref{Vector}, we have already introduced the Wilson loop operator
$\SW_n(C)$ attached to a closed, oriented curve $C$ embedded in $M$, 
\begin{equation}\label{WILSONII}
\SW_n(C) \,=\, \exp{\!\left[i\,n\oint_C A\right]}\,,\qquad\qquad 
n\,\in\,\BZ\,.
\end{equation}
If $C$ is homologically non-trivial, the parameter $n$ must be an
integer to ensure that $\SW_n(C)$ is invariant under ``large,''
homotopically-nontrivial gauge transformations on $M$.  

On the other hand, when $C$ is trivial in $H_1(M;\BZ)$, the condition 
${n\in\BZ}$ can be relaxed.  In the latter case, ${C
  = \partial\Sigma}$ is the boundary of a 
connected, oriented surface ${\Sigma \subset M}$, a so-called Seifert surface
for the knot.  See Ch.\,5 of \cite{Rolfsen:1976} for a nice reference
on Seifert surfaces.  In terms of $\Sigma$, the Wilson loop operator can be
rewritten as 
\begin{equation}\label{WILSONIII}
\SW_\nu(C,[\Sigma]) \,=\, \exp{\!\left[i\,\nu\int_\Sigma
    F_A\right]}\,,\qquad\qquad
\nu\,\in\,\BR\,.
\end{equation}
The expression for $\SW_\nu(C,[\Sigma])$ in \eqref{WILSONIII} is manifestly
gauge-invariant for arbitrary real values of the charge $\nu$, and since $F_A$ is
closed, the operator depends only on the relative homology
class of the Seifert surface,
\begin{equation}
[\Sigma] \,\in\, H_2(M,C)\,.
\end{equation}
The choice of $[\Sigma]$ is an extra discrete choice,
necessary if we wish to extend the definition of the $U(1)$ Wilson loop
operator to non-integral charges.  

As a special case, let us suppose that $M$ is a rational homology
sphere,  with ${b_1 = 0}$ and hence ${H_2(M) =
  0}$.\footnote{For a compact orientable three-manifold, $H_2(M)$ is
  torsion-free.  Thus vanishing of ${b_1 = b_2}$ implies the
  vanishing of $H_2(M)$.}  The relative exact sequence below,
\begin{equation}
\cdots\longrightarrow\, H_2(M) \,\longrightarrow\, H_2(M,C)
\,\buildrel\partial_*\over\longrightarrow\, H_1(C)
\,\buildrel\iota_*\over\longrightarrow\, H_1(M)\,\longrightarrow
\cdots\,,
\end{equation}
implies ${H_2(M,C) \simeq H_1(C) = \BZ}$.  By assumption, the
image ${\partial_*[\Sigma]}$ generates $H_1(C)$, so
$[\Sigma]$ is uniquely determined once the orientation of $C$ is
fixed.  Thus when $M$ is a rational homology sphere, the choice of
Seifert surface can be omitted from our notation for the fractional
Wilson loop operator, and we simply write
\begin{equation}
\SW_\nu(C) \,=\, \exp{\!\left[i\,\nu\int_\Sigma F_A\right]}\,,\qquad\qquad
b_1 = 0\,.
\end{equation}

Both the vertex operator $\SV_k(p)$ and the Wilson loop operator
$\SW_n(C)$ depend upon the particular choices for the point ${p \in M}$
and the curve ${C \subset M}$.  By contrast, the third operator
$\SL_\alpha(C)$ will be homological, depending only upon the class
${[C] \in H_1(M)}$ of the closed curve.  In terms of the periodic
scalar field $\phi$,
\begin{equation}\label{LOPPPHI}
\SL_{\alpha}(C) \,=\, \exp{\!\left[\frac{i\alpha}{2\pi}\oint_C
    d\phi\right]}\,,\qquad\qquad \alpha\,\in\,\BR/2\pi\BZ\,.
\end{equation}
Because the periods of the one-form $d\phi$ are quantized in integral
multiples of $2\pi$, the expression on the right in \eqref{LOPPPHI} is
invariant under a shift ${\alpha \mapsto \alpha + 2\pi}$.  For this
reason, $\alpha$ is best regarded as an angular parameter for the
homological loop operator.

\medskip\noindent{\sl Vertex Operators and Monopoles}\medskip

So far we have introduced three kinds of operators on $M$,
\begin{equation}
\SV_k(p)\,,\qquad\qquad \SW_n(C)\,,\qquad\qquad \SL_\alpha(C)\,,
\end{equation}
labelled generally by parameters 
\begin{equation}
k,\, n\,\in\,\BZ\,,\qquad\qquad \alpha \,\in\,\BR/2\pi\BZ\,.
\end{equation}
The vertex operator $\SV_k(p)$ and the homological loop operator
$\SL_\alpha(C)$ are respectively specified in \eqref{VERTEXP} and
\eqref{LOPPPHI} as classical functionals of the scalar field $\phi$,
whereas the Wilson loop operator $\SW_n(C)$ is a classical functional
of the Maxwell gauge field $A$.

Duality between the scalar and the Maxwell field theories on $M$
implies not only a relation between partition functions, but also a
correspondence between operators in each theory.  So how do we
describe the vertex operator $\SV_k(p)$ and the loop operator
$\SL_\alpha(C)$ dually in the language of Maxwell theory?  And how do we
describe the Wilson loop operator $\SW_n(C)$ in terms of the periodic scalar
field?

In answer to all three questions, the duals of $\SV_k(p)$,
$\SW_n(C)$, and $\SL_\alpha(C)$ will be operators of disorder-type
\cite{tHooft:1977hy}, which create singularities in the dual classical
field.  To quickly explain both the notion of a disorder operator and
its relevance for duality, let us derive the dual of the vertex
operator $\SV_k(p)$.

At first glance, one might be tempted to consider the (unnormalized)
expectation value
\begin{equation}
\big\langle\SV_k(p)\big\rangle \,=\, \int_{\CX}\CD\phi \,\, \SV_k(p)
\, \exp{\!\left[-{\bf I}_{\rm tot}(\phi)\right]}\,.
\end{equation}
Unless ${k=0}$, in which case $\SV_k(p)$ is the identity operator,
$\SV_k(p)$ transforms with charge $k$ under the global $U(1)$ symmetry
in \eqref{GLBLU}.  Hence trivially
\begin{equation}
\big\langle\SV_k(p)\big\rangle \,=\, 0\,,\qquad\qquad k \,\neq\, 0\,,
\end{equation}
due to cancellations in the integral over the constant mode of
$\phi$.  So we cannot learn much by thinking about the expectation
value of $\SV_k(p)$.

Instead, to discuss a non-trivial expectation value, we pick distinct points ${p
  \neq q}$ in $M$ and consider vertex operators of opposite charge
inserted at these points,
\begin{equation}\label{TWOPT}
\big\langle\SV_k(p)\,\SV_{-k}(q)\big\rangle \,=\, \int_{\CX}\CD\phi
\,\, \SV_k(p) \, \SV_{-k}(q) \, \exp{\!\left[-{\bf I}_{\rm
      tot}(\phi)\right]}\,.
\end{equation}
Because the expectation value in \eqref{TWOPT} is invariant under the
global $U(1)$ symmetry, the expectation value need not vanish, and we
can meaningfully ask for the dual description of \eqref{TWOPT} in
terms of the Maxwell gauge field $A$.

Just as for the analysis in Section \ref{DUALPIP}, the first step in
dualizing the vertex operator path integral will be to promote the
integrand in  \eqref{TWOPT} to a functional of the pair
$\left(\phi,B\right)$ which is invariant under the gauge
transformation in \eqref{GAUGEB}.  Since the 
vertex operators $\SV_k(p)$ and $\SV_{-k}(q)$ carry opposite charges,
local gauge invariance can be achieved by introducing a Wilson line
for the auxiliary gauge field $B$ which runs between the vertex operators.  Thus we
choose an oriented curve $\Gamma$ from $q$ to $p$,
\begin{equation}
\partial\Gamma = p \,-\, q\,,\qquad\qquad \Gamma \subset M\,,
\end{equation}
and we consider the expectation value of the gauge-invariant
composite\footnote{We implicitly absorb the
  topological sums over the winding-number $\omega$ and the Chern
  class $c_1(P)$ into the definitions of the spaces ${\CX = \bigsqcup_\omega
    \CX_\omega}$ and ${\CA = \bigsqcup_{c_1(P)} \CA(P)}$.}
\begin{equation}\label{VSPATHI}
\begin{aligned}
&\left\langle\SV_k(p) \, \exp{\!\left[i\,k\int_\Gamma B\right]} \,
\SV_{-k}(q)\right\rangle \,=\,\frac{1}{\Vol\!\left(\CG_A\right)}\frac{1}{\Vol\!\left(\CG_B\right)}\int_{\CX\times\CA\times\CA(P_0)}\mskip
-20mu \CD \phi\,\CD A\,\CD B \,\,\times\\
&\qquad\qquad\qquad\qquad\times\,\exp{\!\left[i\,k \left(\phi(p) - \phi(q)\right) + i \,k \int_\Gamma B \,-\, {\bf I}_{\rm tot}(\phi,A,B)\right]}\,,
\end{aligned}
\end{equation}
evaluated in the full theory of all three fields $\left(\phi,A,B\right)$
with the classical action ${\bf I}_{\rm tot}(\phi,A,B)$ in \eqref{INOAB}.

The Maxwell gauge field $A$ still enters the integrand of \eqref{VSPATHI}
linearly through the Chern-Simons pairing $\SC\SS(A,B)$.  Thus the
path integral over $A$ again produces a delta-function for $B$ with
support on gauge-trivial field configurations.  After we integrate
over $B$ using the delta-function, the extended path integral in
\eqref{VSPATHI} reduces to the path integral over $\phi$ alone in
\eqref{TWOPT},
\begin{equation}
\big\langle\SV_k(p) \,\SV_{-k}(q)\big\rangle \,=\,
\left\langle\SV_k(p) \, \exp{\!\left[i\,k\int_\Gamma B\right]} \,
  \SV_{-k}(q)\right\rangle\,. 
\end{equation}
As a corollary, the extended path integral in \eqref{VSPATHI} does not
depend upon the choice of the curve $\Gamma$ from $q$ to $p$.

Mimicking our previous analysis of the partition function, we
alternately evaluate the path integral in \eqref{VSPATHI} by using the
local action of $\CG_B$ to set ${\phi = 0}$, after which \eqref{VSPATHI} reduces to a path integral
involving only the gauge fields $A$ and $B$,
\begin{equation}\label{VSPATHII}
\begin{aligned}
&\left\langle\SV_k(p) \, \exp{\!\left[i\,k\int_\Gamma B\right]} \,
\SV_{-k}(q)\right\rangle \,=\,\\
&\qquad\qquad\frac{1}{\Vol\!\left(\CG_A\right)} \int_{\CA \times \CA(P_0)} \mskip -10mu \CD A\,\CD
B\,\exp{\!\left[i\,k\int_\Gamma B \,-\, {\bf I}_{\rm
      tot}(A,B)\right]},
\end{aligned}
\end{equation}
where as in \eqref{EFFA},
\begin{equation}
{\bf I}_{\rm tot}(A,B) \,=\, \frac{e^2}{4\pi}\left(B,B\right) \,+\,
\frac{1}{2\pi i}\left\langle F_A, B\right\rangle\,.
\end{equation}

To evaluate the Gaussian integral in \eqref{VSPATHII} further, we
introduce a two-form
$\delta_\Gamma$ which has delta-function support along 
$\Gamma$ and which represents the Poincar\'e dual of the curve,
\begin{equation}\label{CURRENT}
\exp{\!\left[i\,k\int_\Gamma B\right]} \,=\, \exp{\!\left[i\,k\int_M
    \delta_\Gamma\^ B\right]}\,,\qquad\qquad \delta_\Gamma\,\in\,\Omega^2(M)\,,
\end{equation}
so that all terms in the argument of the exponential take the form of
integrals over $M$.  Because $\Gamma$ is bounded by the points $p$ and
$q$, the two-form $\delta_\Gamma$ is not closed but rather satisfies
the distributional identity
\begin{equation}\label{STOKES}
d\delta_\Gamma \,=\, -\delta_p \,+\, \delta_q\,,\qquad\qquad \delta_p, \delta_q
\,\in\, \Omega^3(M)\,.
\end{equation}
By definition, $\delta_p$ and $\delta_q$ are three-forms with delta-function
support at the points $p$ and $q$.  The identity in \eqref{STOKES} is
most easily deduced as a consequence of Stokes' theorem for the path
$\Gamma$.  For if ${f\in\Omega^0(M)}$ is any smooth function on $M$, then
\begin{equation}
\begin{aligned}
f(p) \,-\, f(q) \,=\, \int_\Gamma df \,=\, \int_M \delta_\Gamma\^df
\,=\, -\int_M d\delta_\Gamma \cdot f\,.
\end{aligned}
\end{equation}
See for instance Chapter $3$ of \cite{Griffiths:78} for more about
distributional differential forms like our $\delta_\Gamma$.

Via the definition in \eqref{CURRENT}, the Gaussian integral over
$A$ and $B$ takes precisely the same form as the generalized partition
function in \eqref{GENZPTHR} with singular ${\upgamma = 2\pi k \,
  \delta_\Gamma}$,
\begin{equation}\label{SOMIAB}
\begin{aligned}
&\left\langle\SV_k(p) \, \exp{\!\left[i\,k\int_\Gamma B\right]} \,
\SV_{-k}(q)\right\rangle \,=\,\\
&\qquad\qquad\frac{1}{\Vol\!\left(\CG_A\right)} \int_{\CA \times \CA(P_0)} \mskip -10mu \CD A\,\CD
B\,\exp{\!\left[-\frac{e^2}{4\pi}\left(B,B\right) +
    \frac{i}{2\pi}\left\langle F_A + 2\pi k \,\delta_\Gamma, B\right\rangle\right]}.
\end{aligned}
\end{equation}
After performing the path integral over $B$, we obtain the desired
reformulation 
\begin{equation}\label{MODMAX}
\big\langle\SV_k(p)\,\SV_{-k}(q)\big\rangle \,=\,
\frac{1}{\Vol\!\left(\CG_A\right)} \int_\CA\CD A\,
\exp{\!\left[-\frac{1}{4\pi
      e^2}\left(\CF_A,\,\CF_A\right)\right]}\,,
\end{equation}
where 
\begin{equation}
 \CF_A \,=\, F_A
\,+\, 2\pi k \, \delta_\Gamma\,.
\end{equation}

The interpretation of the modified Maxwell path integral in
\eqref{MODMAX} is by now well understood.  Due to the explicit delta-function in
$\CF_A$, the argument of the exponential diverges (and thus the
integrand vanishes) unless $F_A$ itself has 
the appropriate singularity along $\Gamma$ to cancel the
delta-function in $\CF_A$,
\begin{equation}\label{SINGD}
F_A \,=\, -2\pi k \, \delta_\Gamma \,+\, \cdots\,,
\end{equation}
where the ellipses indicate regular terms in $F_A$.  Thus, the
insertion of the vertex operators $\SV_k(p)$ and $\SV_{-k}(q)$ in the
scalar sigma model is interpreted dually as the instruction to perform
the Maxwell path integral over connections with the specified singular
behavior along $\Gamma$.  Operators defined in this manner, as an
instruction to perform the path integral over fields with given 
classical singularities, are said to be of disorder-type.

Because $\delta_\Gamma$ is not closed, the Bianchi identity for $F_A$
is modified by the singularity in \eqref{SINGD},
\begin{equation}\label{MONOP}
d F_A \,=\, 2 \pi k \left(\delta_p \,-\, \delta_q\right),
\end{equation}
where we apply the Stokes' identity in \eqref{STOKES}.  Physically,
the new source terms in the Bianchi identity for $F_A$ are interpreted as
magnetic monopoles of charges $\pm k$ at the points $p$ and $q$.
Otherwise, so long as $k$ is integral, the Dirac string singularity
along the curve $\Gamma$ is a gauge artifact. 

In light of \eqref{MONOP}, we see that the operator $\SV_k(p)$
itself is the monopole operator of charge $k$ in the abelian gauge
theory.  By definition, the monopole operator of magnetic
charge $k$ is the local disorder operator which creates a curvature
singularity in $A$ at $p$ of the form 
\begin{equation}\label{MONOPII}
F_A \,=\, -\frac{k}{2}\,\*d\!\left(\frac{1}{r}\right)\,,
\end{equation}
where $r$ is a local radial coordinate centered at $p$.  With this
singularity, the integral of $F_A$ over any small sphere centered
about $p$ is equal to $2\pi k$, as required by the Bianchi identity in
\eqref{MONOP}.  Also, with the given local behavior in \eqref{MONOPII}, $F_A$
satisfies the classical source-free Maxwell equation ${d\*F_A = 0}$
on a punctured neighborhood of the point $p$.

\medskip\noindent{\sl Vortex Loops and Wilson Loops}\medskip

The loop operators $\SL_\alpha(C)$ and $\SW_n(C)$ can be followed
through the duality in much the same fashion as the vertex operator
$\SV_k(p)$.  Very briefly, to dualize the homological loop operator
$\SL_\alpha(C)$ in \eqref{LOPPPHI}, we consider its gauge-invariant
extension in terms of the pair $(\phi, B)$,
\begin{equation}\label{DEFLP}
\SL_\alpha(C) \,=\, \exp{\!\left[\frac{i\,\alpha}{2\pi}\oint_C
    d_B\phi\right]} \,=\, \exp{\!\left[\frac{i\,\alpha}{2\pi}\int_M
    \delta_C \^\!\left(d\phi + B\right)\right]}\,.
\end{equation}
Here $\delta_C$ is a two-form with delta-function support which
represents the Poincar\'e dual of the closed curve ${C \subset M}$.

If we consider the expectation value of $\SL_\alpha(C)$ in
the extended theory of triples $\left(\phi, A, B\right)$ with total
action \eqref{INOAB}, the path integral over $A$ still provides a
delta-function with support on gauge-trivial configurations for $B$.
With this delta-function, $B$ can then be gauged to zero to recover
the expectation value for $\SL_\alpha(C)$ in the theory of the
periodic scalar field $\phi$ alone.

Alternatively, $\phi$ can be gauged to zero in the extended theory of
triples $\left(\phi, A, B\right)$, after which we encounter a Gaussian
integral over $B$  taking precisely the same form as \eqref{SOMIAB}.  Hence
the operator $\SL_\alpha(C)$ is interpreted in the dual Maxwell
theory as a disorder operator which creates a curvature singularity along $C$, 
\begin{equation}\label{SURFACO}
F_A \,=\, -\alpha \, \delta_C \,+\, \cdots\,.
\end{equation}
This curvature singularity looks very much like the preceding 
singularity \eqref{SINGD} which we interpreted in terms of monopoles.
However, $C$ is now closed, without boundary, and $\alpha$ is not a
multiple of $2\pi$.  As a result, the physical interpretation of
\eqref{SURFACO} is different.

In a small tubular neighborhood of $C$, the singularity in \eqref{SURFACO} implies that the gauge
field $A$ behaves as 
\begin{equation}\label{MONO}
A \,=\, -\frac{\alpha}{2\pi}\,d\vartheta \,+\, \cdots\,,
\end{equation}
where $\vartheta$ is an angular coordinate on the plane transverse to
$C$, located at the origin.  By Stokes' theorem, the angular form $d\vartheta$
satisfies ${d(d\vartheta) = 2\pi \delta_C}$, from which \eqref{SURFACO}
follows.  

Evidently, in the presence of the loop operator $\SL_\alpha(C)$, the
gauge field $A$ has non-trivial monodromy ${\Lambda =
  \exp{\!\left(-i\,\alpha\right)}}$ 
about any small curve which links $C$.  Of course, the value of the
monodromy only depends upon the value of $\alpha$ modulo $2\pi$.
Physically, shifts in $\alpha$ by units of $2\pi$ can be accomplished
by gauge transformations ${u:M\to U(1)}$ which are themselves singular
along $C$, of the local form 
\begin{equation}\label{SINGG}
u \,=\, \e{i n \vartheta}\,,\qquad\qquad n \,\in\,\BZ\,.
\end{equation}
When $\alpha$ in \eqref{MONO} is an integral multiple of $2\pi$, the singularity
in $A$ can be removed by such a gauge transformation, but not otherwise.

For gauge theories in four dimensions, the codimension-two
singularity in \eqref{MONO} defines the basic Gukov-Witten
\cite{Gukov:2006jk} surface operator.  Hence the loop operator
$\SL_\alpha(C)$ in three dimensions can be interpreted as the
reduction of a surface operator from four dimensions.  From the purely
three-dimensional perspective, $\SL_\alpha(C)$ can be considered as a
kind of monodromy or vortex loop.

Reversing directions, we finally discuss the interpretation for the Wilson loop
operator $\SW_n(C)$ in terms of the periodic scalar field $\phi$.  For the time
being, we do not make any assumption about the homology class of $C$,
so the charge ${n\in\BZ}$ must be an integer to maintain invariance
under arbitrary gauge transformations.  

In terms of the two-form $\delta_C$ with delta-function support, the
abelian Wilson loop operator can be rewritten as  
\begin{equation}
\SW_n(C) \,=\, \exp{\!\left[i\,n\oint_C A\right]} \,=\,
\exp{\!\left[i\,n\int_M\delta_C\^A\right]}\,.
\end{equation}
The expectation value of $\SW_n(C)$ can now be evaluated in the
extended theory of triples $(\phi,A,B)$,
\begin{equation}\label{WLOPPH}
\begin{aligned}
&\big\langle\SW_n(C)\big\rangle \,=\, \frac{1}{\Vol\!\left(\CG_A\right)}\frac{1}{\Vol\!\left(\CG_B\right)}\int_{\CX\times\CA\times\CA(P_0)}\mskip
-20mu \CD \phi\,\CD A\,\CD B\,\,\times\\
&\qquad\qquad\qquad\times
\exp{\!\left[-\frac{e^2}{4\pi}\left(d_B\phi,d_B\phi\right) + 
    \frac{i}{2\pi}\left\langle F_A,B\right\rangle + i \, n
    \left\langle\delta_C, A\right\rangle\right]}\,.
\end{aligned}
\end{equation}
By gauging $\phi$ to zero and performing the Gaussian integral over
$B$, one sees that the extended path integral in \eqref{WLOPPH}
describes the usual Wilson loop expectation value in Maxwell theory on $M$.  

On the other hand, as also clear from \eqref{WLOPPH}, the gauge field $A$
still enters the argument of the exponential linearly.  Due to the
new term involving $\delta_C$, the path integral over $A$ now
produces a delta-function for $B$ that enforces the condition 
\begin{equation}\label{CURVB}
F_B \,=\, 2\pi n \, \delta_C\,,
\end{equation}
and $B$ has trivial holonomy otherwise.  

At first glance, one might think that the curvature condition on $B$
is vacuous, since we have already noted, in the discussion of the corresponding
singularity for $A$, that the singularity in \eqref{CURVB} can be 
removed by a gauge transformation of the local form in \eqref{SINGG}.
However, we must remember that the group $\CG_B$ acts
simultaneously on both $B$ and $\phi$ via \eqref{GAUGEB}, so if we
perform a gauge transformation to remove the singularity in $B$, we will
create a singularity in $\phi$!

Specifically, once we perform the Wilson loop path integral over $A$
in \eqref{WLOPPH} and select a representative for the connection $B$ 
satisfying \eqref{CURVB}, with trivial holonomies otherwise, we can
rewrite the Wilson loop expectation value strictly in terms of $\phi$,
\begin{equation}\label{WLLOPHII}
\big\langle\SW_n(C)\big\rangle \,=\, \int_\CX \CD\phi \,\,
\exp{\!\left[-\frac{e^2}{4\pi}\left(d_B\phi,
      d_B\phi\right)\right]}\,,\qquad\qquad d_B \phi \equiv d\phi + B\,.
\end{equation}
Here $B$ is now a background, spectator field, and we have fixed the
action of $\CG_B$ with our choice of representative connection.  

If we wish to eliminate $B$ entirely, we can introduce
a new periodic scalar field $\wt\phi$, defined so that
\begin{equation}
d\wt\phi \,=\, d\phi + B\,,
\end{equation}
after which
\begin{equation}
\big\langle\SW_n(C)\big\rangle \,=\, \int_\CX \CD\wt\phi \,\,
\exp{\!\left[-\frac{e^2}{4\pi}(d\wt\phi,
      d\wt\phi)\right]}\,.
\end{equation}
Similar to \eqref{MONO}, the background connection $B$ behaves in a
neighborhood of $C$ as ${B = n \, d\vartheta}$.  Thus, $\wt\phi$ must be
related to $\phi$ near $C$ by 
\begin{equation}\label{NEWPHI}
\wt\phi \,=\, \phi \,+\, n \, \vartheta\,.
\end{equation}
From \eqref{NEWPHI} we see that $\wt\phi$ winds non-trivially 
around any small curve which encircles $C$.
As a result, the Wilson loop operator $\SW_n(C)$ is interpreted dually
as the disorder operator which creates an additive monodromy in $\phi$ 
of $n$ units about the meridian of $C$.  From the dual perspective,
the integrality of the charge $n$ is necessary to ensure that $\phi$ is
single-valued as a map from the knot complement ${M^o = M - C}$ to $S^1$.

When $C$ is trivial in $H_1(M)$, we have noted that
the parameter ${n\in\BZ}$ in the Wilson loop operator can be extended to an
arbitrary real number ${\nu \in \BR}$, as appears in
\eqref{WILSONIII}.  By our preceding discussion, the operator
$\SW_\nu(C)$ then creates a fractional monodromy in $\phi$.  

To understand the fractional monodromy better, let us think about
dualizing the Wilson loop operator in the form
\begin{equation}
\SW_\nu(C) \,=\, \exp{\!\left[i\,\nu\int_\Sigma F_A\right]} \,=\,
\exp{\!\left[i\,\nu\int_M
    \delta_\Sigma\^F_A\right]}\,.
\end{equation}
As before, $\Sigma$ is a Seifert surface bounding $C$.  For
convenience, I assume that $M$ is a rational homology sphere, with
${b_1 = 0}$, so that the relative homology class of $\Sigma$ is
unique.  Associated to $\Sigma$ is the Poincar\'e dual one-form
$\delta_\Sigma$ with delta-function support on $\Sigma$ and satisfying the
distributional identity ${d\delta_\Sigma = \delta_C}$.

By the same observations which we applied to \eqref{WLOPPH},
$\SW_\nu(C)$ is described in terms of $\phi$ as the instruction to perform the
path integral over $\phi$ in \eqref{WLLOPHII} with a background
connection $B$ which now satisfies 
\begin{equation}\label{SINGB}
B \,=\, 2\pi\nu \, \delta_\Sigma\,.
\end{equation}
Equivalently, we replace $\phi$ by a new field $\wt\phi$ so that
${d\wt\phi = d\phi + B}$ with the given $B$.

To characterize the local behavior of $\wt\phi$ near $C$, we assume
that $M$ is $\BR^3$, with coordinates $(x,y,z)$, and that the curve $C$
extends upwards along the $z$-axis.  We then take $\Sigma$ to be the portion
of the $xz$-plane with ${x \ge 0}$.  Hence $y$ is the local coordinate
normal to $\Sigma$.  In these local coordinates, the expression for
$B$ in \eqref{SINGB} just becomes ${B = 2\pi\nu\, H(x) \, \delta(y) \,
  dy}$, where $H(x)$ is the Heaviside step-function.\footnote{By definition
  ${H(x) = 1}$ for ${x > 0}$, and ${H(x) = 0}$ for ${x < 0}$.}  See
Figure 1 for a sketch of the situation.
\begin{figure}[htb!]
\begin{center}
\includegraphics[scale=0.50]{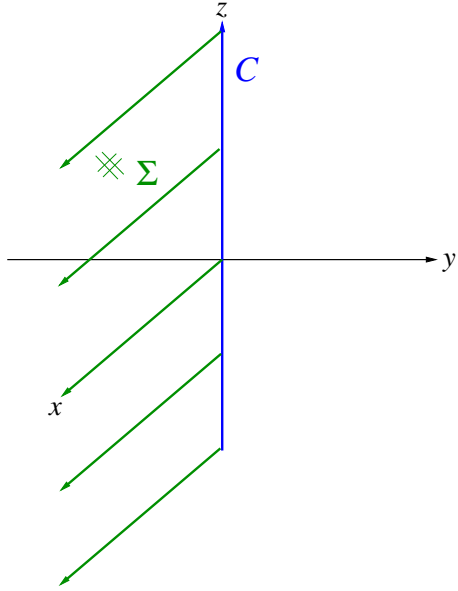}
\caption{Seifert surface $\Sigma$ attached to $C$.}
\end{center}
\end{figure}

To obtain a corresponding local description for $\wt\phi$ near
$\Sigma$, we integrate the relation ${d\wt\phi = d\phi + B}$ using our
expression for $B$.  In the region near $\Sigma$ with ${x > 0}$,
we find that $\phi$ is related to $\wt\phi$ by 
\begin{equation}
\wt\phi \,=\, \phi \,+\, 2 \pi \nu \, H(y)\,,\qquad\qquad x > 0\,.
\end{equation}
The Heaviside function $H(y)$ arises from the integral of the
delta-function $\delta(y)$.  Thus, when $C$ is
null-homologous and the charge ${\nu\in\BR}$ is 
fractional, the role of the Wilson loop operator $\SW_\nu(C)$
is dually to insert a discontinuity in the sigma model field $\phi$
transverse to the Seifert surface $\Sigma$.  Physically, $\Sigma$
can be interpreted as a kind of domain wall which is created by the
fractional Wilson loop operator $\SW_\nu(C)$.

\bigskip
\noindent{\bf Acknowledgments}\smallskip

I take pleasure in thanking Marcus Benna, Martin Ro\v cek, and
A.J.\,Tolland for conversations on these and related matters.
I acknowledge support from the Simons Center for Geometry and Physics,
where a portion of this work was completed.

\end{onehalfspace}
\bibliographystyle{unsrt} 

\end{document}